\documentclass[12pt]{article}

\usepackage[colorlinks = true,
linkcolor = black,
urlcolor  = blue,
citecolor = black,
anchorcolor = black]{hyperref}

\usepackage{lineno}

\usepackage{scicite}

\usepackage{times}

\usepackage{amsmath}
\usepackage{amsfonts}
\usepackage{amssymb}
\usepackage{graphicx}
\usepackage[euler]{textgreek}

\newenvironment{sciabstract}{%
\begin{quote} \bf}
{\end{quote}}

\title{Probing electron and hole co-localization by resonant four-wave mixing in the extreme-ultraviolet}

\author{
Horst Rottke,$^{1\ast}$ Robin Y. Engel,$^{3,4}$ Daniel Schick,$^1$\\
Jan O. Schunck,$^{3,4}$ Piter S. Miedema,$^3$ Martin C. Borchert,$^1$\\
Marion Kuhlmann,$^3$ Nagitha Ekanayake,$^3$ Siarhei Dziarzhytski,$^3$\\
G\"{u}nter Brenner,$^3$ Ulrich Eichmann,$^1$  Clemens von Korff Schmising,$^1$\\
Martin Beye,$^{3,4}$ and Stefan Eisebitt$^{1,2}$\\
\\
\normalsize $^1$Max-Born-Institut f\"{u}r Nichtlineare Optik und Kurzzeitspektroskopie,\\
\normalsize Max-Born-Stra\ss{}e 2A, 12489 Berlin, Germany\\
\normalsize $^2$Technische Universität Berlin, Institut für Optik und Atomare Physik,\\
\normalsize Straße des 17. Juni 135, 10623 Berlin, Germany\\
\normalsize $^3$Deutsches Elektronen Synchrotron DESY, Notkestr. 85, 22607 Hamburg, Germany\\
\normalsize $^4$Physics Department, Universität Hamburg, Luruper Chaussee 149, 22761 Hamburg, Germany\\
\\
\normalsize{$^\ast$Corresponding author. E-mail:  rottke@mbi-berlin.de}
}

\date{}

\begin{document} 

\baselineskip24pt


\maketitle 

\textbf{Short title: Resonant four way mixing with x-rays as a probe of charge localization}\\

\begin{sciabstract}
  The extension of nonlinear spectroscopic techniques into the x-ray domain is in its infancy but holds the promise to provide unique insight into the dynamics of charges in photoexcited processes, which are of fundamental as well as applied interest.  We report on the observation of a third order nonlinear process in lithium fluoride at a free-electron laser. Exploring the yield of four wave mixing (FWM) in resonance with transitions to strongly localized core exciton states vs. delocalized Bloch states, we find resonant FWM to be a sensitive probe for the degree of charge localization: substantial sum- and difference-frequency generation is observed exclusively when in a one- or three-photon resonance with a LiF core exciton, with a dipole forbidden transition affecting details of the nonlinear response.   Our reflection-geometry-based approach to detect FWM signals enables the study of a wide variety of condensed matter sample systems, provides atomic selectivity via resonant transitions and can be easily scaled to shorter wavelengths at free electron x-ray lasers.
\end{sciabstract}

\paragraph*{Introduction}

Nonlinear wave mixing in the \textit{optical} spectral range is a cornerstone of nonlinear optics. It has been used extensively for the generation of coherent light at new, otherwise inaccessible wavelengths, for the analysis of optical properties of materials and, using ultra-short light pulses, to explore the evolution of optical excitations in time.
Free-electron lasers (FELs) have opened the opportunity to apply these nonlinear techniques in the \textit{extreme ultraviolet} (XUV) up to the \textit{x-ray} spectral ranges. One promise of these approaches is that at sufficiently short wavelengths and pulse duration, electronic excitations may be followed on an atomic length and time scale.
Moreover, XUV radiation allows to operate in resonance with core levels providing element specificity and spectroscopic information with little lifetime broadening.

So far, however, there have been only few studies on nonlinear processes in solids involving XUV or x-ray beams.
Parametric down conversion, a second order nonlinear process, aimed at probing valence-electron charge densities and the optical response of crystals with atomic resolution~\cite{Danino1981,Tamasaku2011,Schori2017}.
With similar aims, second order x-ray and optical sum-frequency mixing, essentially being optically modulated x-ray diffraction, has first been observed using FEL and long wavelength optical laser radiation~\cite{Glover2012}.
Proof-of-principle experiments demonstrating  second harmonic generation of x-rays in solids  and interface sensitivity of
second order nonlinear processes in centrosymmetric crystals in the soft x-ray range  have also been reported~\cite{Shwartz2014,Lam2018}.

In contrast, four wave mixing (FWM), as a third order nonlinear process, is able to address the bulk properties of any single crystal structure, irrespective of its symmetry.  Here, first experimental investigations based on transient gratings induced by XUV FEL pulses and probed either by optical or even XUV laser pulses have been reported~\cite{Bencivenga2015,Foglia2018,Bencivenga2019}. Very recently, FWM in NaCl at the $\mathrm{Na}^+$~$L_{2,3}$ edge near 33.5\,eV, using laser-generated attosecond XUV and near infrared (NIR) laser pulses,  revealed the presence of dark excited states, i.e. states not dipole coupled to the electronic ground state of the material~\cite{Gaynor2021}.

We report on the observation of a complementary FWM process driven in a solid, namely sum- and difference-frequency mixing involving one XUV ($\omega_\text{X}$) and two NIR photons ($\omega_\text{I}$) with resulting beams at frequencies $\omega_\text{FWM} = \omega_\text{X} \pm 2\omega_\text{I}$, i.e. different from the radiation driving the process, which to our knowledge has not been observed yet. 
We study LiF, a prototypical ionic crystal with centrosymmetry. 
In particular, we investigate the effect of electronic excitations with a high degree of electron-hole correlation on these processes, utilizing the presence of both stronglgy localized core excitons and delocalized Bloch states in LiF as prototypical model to probe the degree of co-localization of the excited electron and hole.

Exploiting the wavelength tunability of the FEL source we are able to chart the FWM conversion efficiency. We find significant sum- and difference-frequency generation in LiF when in either one- or three-photon resonance with the core exciton which is linked with the Li$^+$ $1s2p$ excited electron configuration. An energetically near exciton, linked with the Li$^+$ $1s2s$ electron configuration and suspected to be present from linear absorption experiments, is seen - via comparison to theory - to affect the frequency conversion processes via two-photon resonance and a near resonant one-NIR-photon coupling of both excitonic states. In contrast, we do not observe any FWM when exclusively in resonance with transitions to delocalized Bloch states in the experiment.
This indicates that FWM in the XUV and soft x-ray spectral range is a sensitive probe of charge co-localization and charge separation.
Given that the separation of holes and electrons and the localization of charge at particular atomic sites are key steps in (photo)chemical reactions, light harvesting and the elementary steps of photovoltaics, we expect this sensitivity to be instrumental in the study of phenomena at the interface of materials science, physics and chemistry in the future.

\paragraph*{Results}

\begin{figure}[h]
\centering

\includegraphics[width=0.6\textwidth]{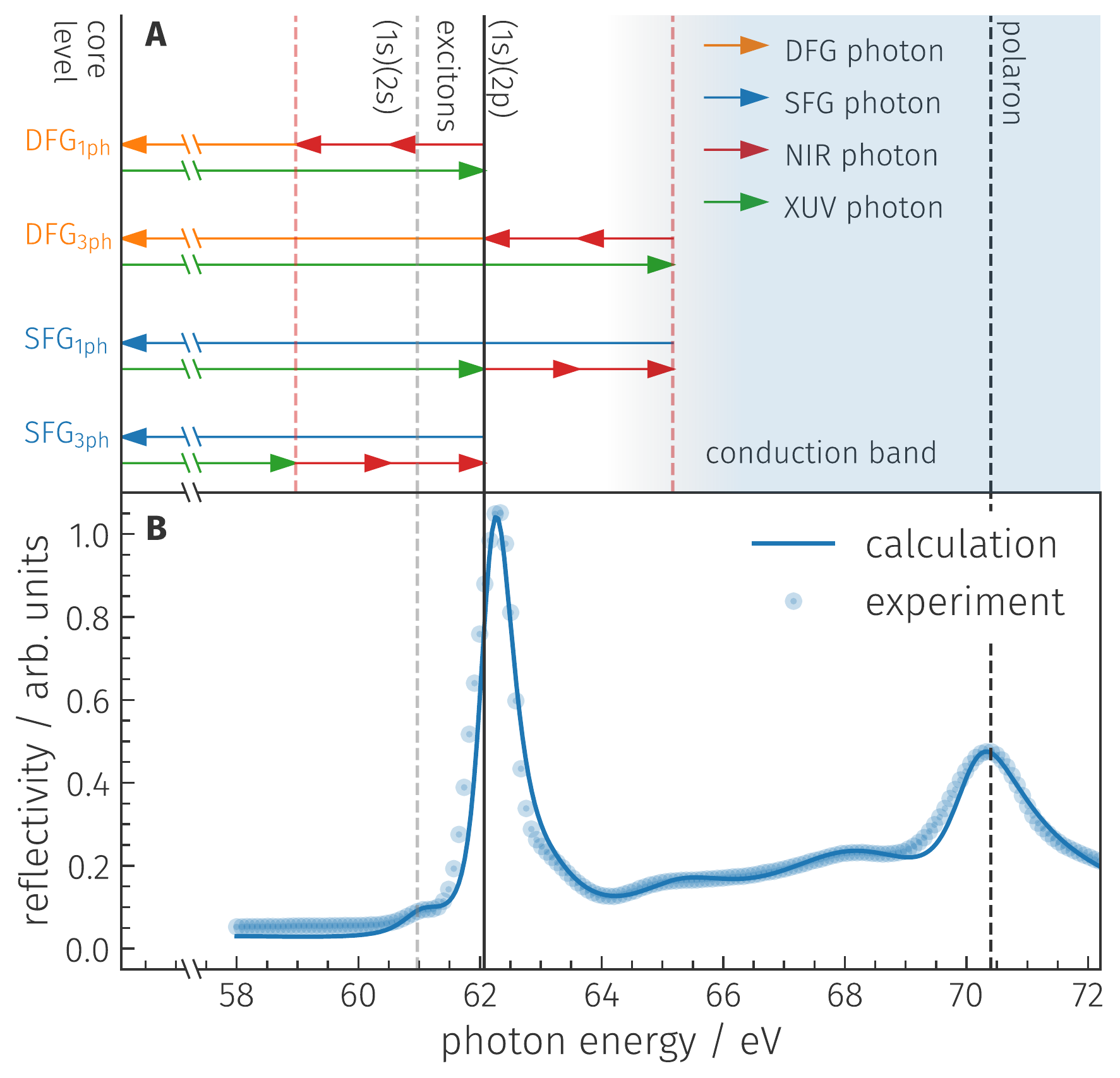}

\caption{
	\textbf{Four wave mixing schemes and the LiF linear reflectivity.}
	\textbf{(A)} Schematic view of the LiF level structure relevant to the experiment with the horizontal arrows representing sum- and difference-frequency mixing in one- and three-photon resonance with the $p$-type LiF core exciton at 62.07\,eV photon energy, just below the conduction band edge.
	The green and red arrows depict the driving XUV and NIR photons, respectively. 
	The blue and orange ones represent the generated sum- and difference-frequency photons, respectively.
	\textbf{(B)} The schematic level structure of \textbf{(A)} showing up in the linear reflectivity of LiF in the photon energy range relevant to the experiment.
	The blue dots represent the measured reflectivity.
	The line is the calculated reflectivity based on the model for the linear dielectric constant $\epsilon$ as outlined in the supplement.
	The narrow reflectivity peak visible at 62.3\,eV is due to the LiF $p$-type core exciton resonance located at 62.07\,eV (vertical line) with one of the $\mathrm{Li^+\, (1s)^2}$ core electrons excited.
	Conduction band absorption is generally assumed to start at photon energies to the right of this resonance.}

\label{processes}

\end{figure}

The frequency mixing processes studied, one- and three-photon resonant with a LiF core exciton, are schematically illustrated in Fig.~\ref{processes}(A).
The core exciton of interest here is formed by the promotion of one of the strongly localized $1s$ core electrons of a $\mathrm{Li}^+$ ion in LiF to an excited state equally localized on the same ion and linked with the $\mathrm{Li}^+$ electronic configuration $1s2p$.
It gives rise to a dipole allowed transition with a characteristic narrow absorption structure in linear spectroscopy~\cite{Haensel1968}.
We will refer to this final state as  \emph{p-type} exciton.
The prominent, narrow resonance enhancement of the LiF reflectivity at 62.3\,eV photon energy in Fig.~\ref{processes}(B) is attributable to this core exciton.
The $1s2s$ excited electron configuration  of the $\mathrm{Li}^+$ ion is energetically close by, but  not dipole coupled to the $(1s)^2$ ground state configuration in a perfect cubic LiF crystal.
We will refer to this $1s2s$ $\mathrm{Li}^+$ electron configuration as \emph{s-type} exciton.
In linear absorption spectroscopy~\cite{Haensel1968} and in the LiF linear reflectivity shown in Fig.~\ref{processes}(B) a small low energy shoulder of the $p$-type exciton feature, located at $\approx 61$\,eV, is suspected to be due to this $s$-type exciton, becoming weakly visible via dipole coupling to the $(1s)^2$ configuration through inversion symmetry breaking~\cite{Olovsson2009a}.
In contrast, transitions into delocalized band states are responsible for the spectral features at higher photon energies in Fig.~\ref{processes}(B)~\cite{Pantelides1975,Fields1977,Shirley2004,Olovsson2009}.

\begin{figure}[h]
\centering
\includegraphics[width=0.75\textwidth]{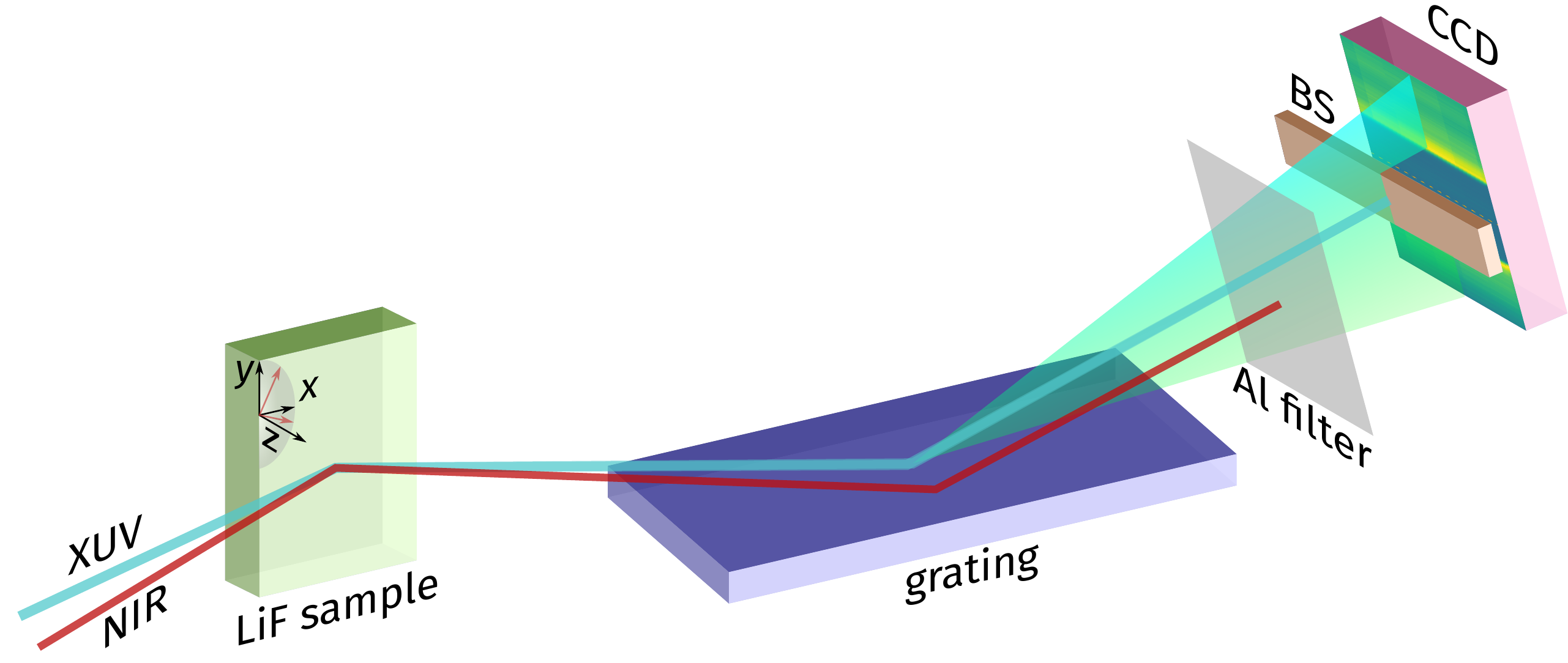}
\caption{
	\textbf{Schematic view of the experimental setup.}
	The FEL (blue), NIR (red) and the generated sum- and difference-frequency beams reflected off the LiF to vacuum interface are dispersed by a reflection grating. 
	Subsequently, the NIR beam is blocked by an aluminum foil and the FEL beam by a separate beam stop (BS) directly in front of a CCD camera which detects the generated sum- and difference frequency beams. 
	The XUV and NIR beams propagate in a common plane of incidence towards the LiF sample which we choose to be the $x,z$-plane of a suitably chosen Cartesian coordinate system with the $z$-axis normal to the LiF surface and parallel to one of the LiF cubic crystal axes (see the axis system indicated on the LiF crystal surface). 
	The two in plane axes of the LiF crystal (red axes) were rotated by $(22.5 \pm 2.5)^\circ$ with respect to the $x,y$-axes of this coordinate system in the LiF surface plane.}
\label{setup}
\label{geometry}
\end{figure}

We study FWM at the sum- and difference-frequencies $\omega_\mathrm{X}\, \pm \, 2\omega_\mathrm{I}$ by detecting the generated radiation emitted by a LiF single crystal towards the vacuum side, from where the process driving XUV and NIR laser beams are impinging.
The scheme of the setup in Fig.~\ref{setup} visualizes this detection approach;
the experimental parameters are detailed in the methods section.

\begin{figure}[h]
\centering
\includegraphics[width=0.65\textwidth]{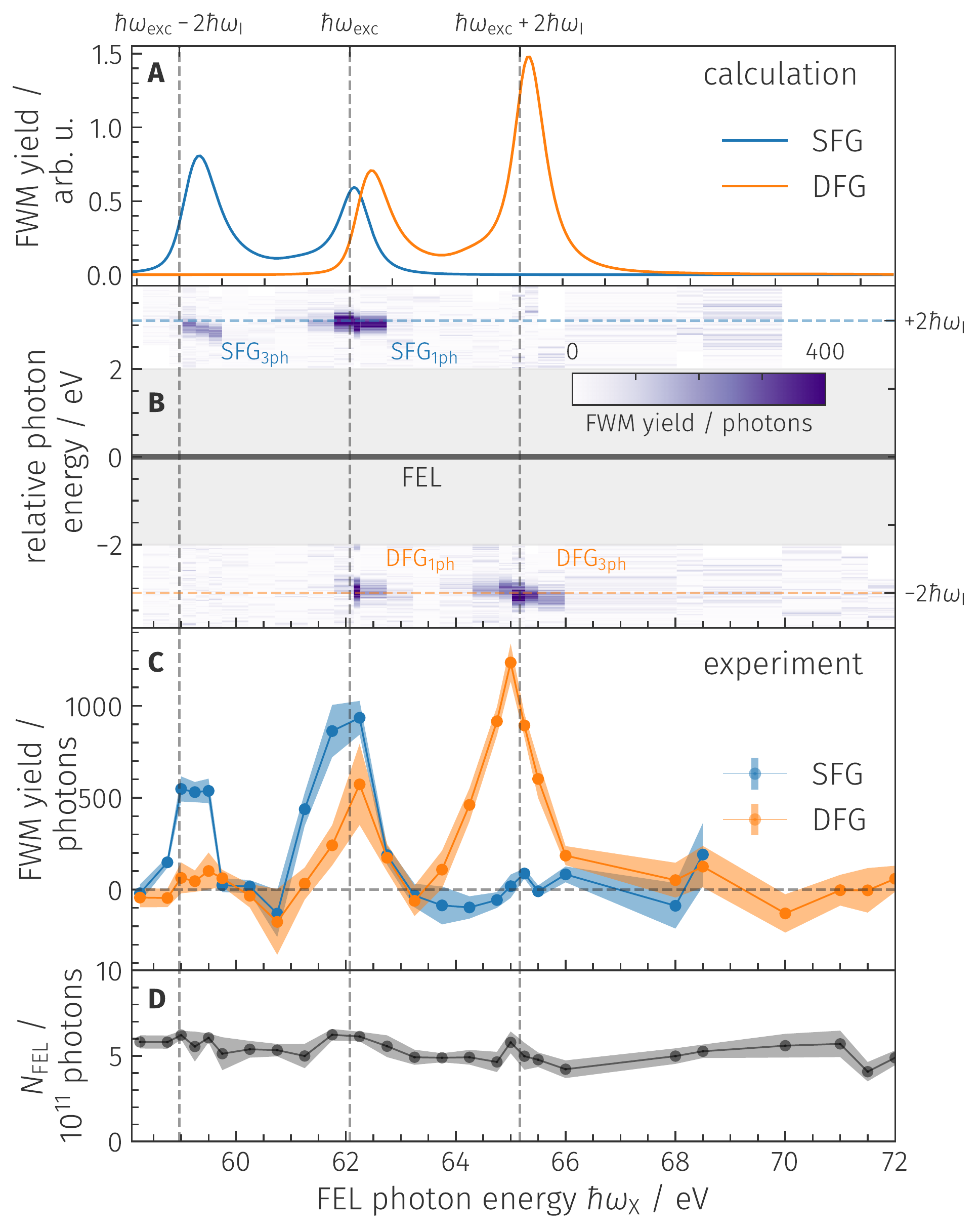}
\caption{\textbf{Measured and calculated frequency conversion yields.}
    \textbf{(A)} Calculated sum- and difference frequency yields plotted as functions of the process driving FEL photon energy. The NIR photon energy is fixed at 1.55\,eV.
	The plot is based on our basic model of the LiF third order nonlinear susceptibility tensor in this photon energy range and corresponds to the parameter $R=0.112$ in Fig.~\ref{calc}.
	\textbf{(B)} Density plot of the experimental spectral distributions of the generated sum- and difference frequency photons (vertical axis) for the different FEL photon energy settings (horizontal axis).
	The spectral distributions are shown relative to the respective FEL photon energy settings. They thus appear at an offset of $\pm 2\hbar\omega_\text{I} = \pm3.1\, \mathrm{eV}$ with respect to the respective FEL photon energy (horizontal line at zero).
	\textbf{(C)} The generated number of sum- (blue) and difference-frequency (orange) photons plotted over the process driving FEL photon energy $\hbar\omega_\text{X}$.
	Shown is the total number of generated photons per FEL pulse-train at zero FEL-NIR pulse delay summed over the whole spectral distribution of the generated radiation.  Sizeable frequency conversion is observed only with the incident FEL and NIR radiation in one- or three-photon resonance with the $p$-type core exciton of LiF.
	\textbf{(D)} The number of FEL photons per pulse-train incident on the LiF sample at the FEL photon energies in \textbf{(C)}.
	Shaded areas in panels \textbf{(C)} and \textbf{(D)} connect 95\% confidence intervals of the data.
}
\label{fig2}
\end{figure}

In Fig.~\ref{fig2} we show our main experimental result, the dependence of the sum- and difference-frequency conversion yields on the setting of the FEL photon energy in a range covering the LiF core exciton and part of the conduction band.
The density plot in Fig.~\ref{fig2}(B) reveals the energy offset of the  radiation generated by wave mixing with respect to the corresponding setting of the FEL photon energy.
As directly evident from the experimental data, it amounts to $\pm 2\hbar\omega_\text{I} = \pm 3.1\,$eV, i.e. two times the fixed NIR photon energy, for the generated sum- and difference-frequency with narrow spectral distributions.
This spectral spacing is indicative of the wave-mixing nature of the signals observed. Fig.~\ref{fig2}C presents individually the total number of sum- and difference-frequency photons generated in each FEL pulse train during XUV and NIR temporal overlap as function of the photon energy of the process driving XUV radiation.
The corresponding number of XUV photons from the FEL impinging on the LiF sample per pulse train for each measuring point in (B,~C) are shown in Fig. \ref{fig2}(D).
The direction of polarization of both, the FEL and NIR laser pulses, is in the common plane of incidence on the sample and both pulses arrive simultaneously except for a statistically distributed relative arrival time jitter.
The entire measurement is performed at a fixed NIR pulse intensity on the LiF crystal.
As is obvious in Figs.~\ref{fig2}(B,~C), we observe a significant yield of frequency conversion only when in resonance with the $p$-type core exciton located at $\hbar \omega_\text{exc} = 62.07\,$eV. This can either be a one photon resonance with the FEL photon energy tuned to resonance ($\omega_\text{X} \approx \omega_\text{exc}$), or a three photon resonance satisfying $\omega_\text{X}\, \pm \, 2\omega_\text{I} \approx \omega_\text{exc}$. 
In both cases, sum- and difference-frequency generation is observed as sketched in Fig.~\ref{processes}(A).
Aside from these settings, no FWM-frequency conversion was detectable above the residual background level of FEL stray light.

\begin{figure}[h]
\centering
\includegraphics[width=0.75\textwidth]{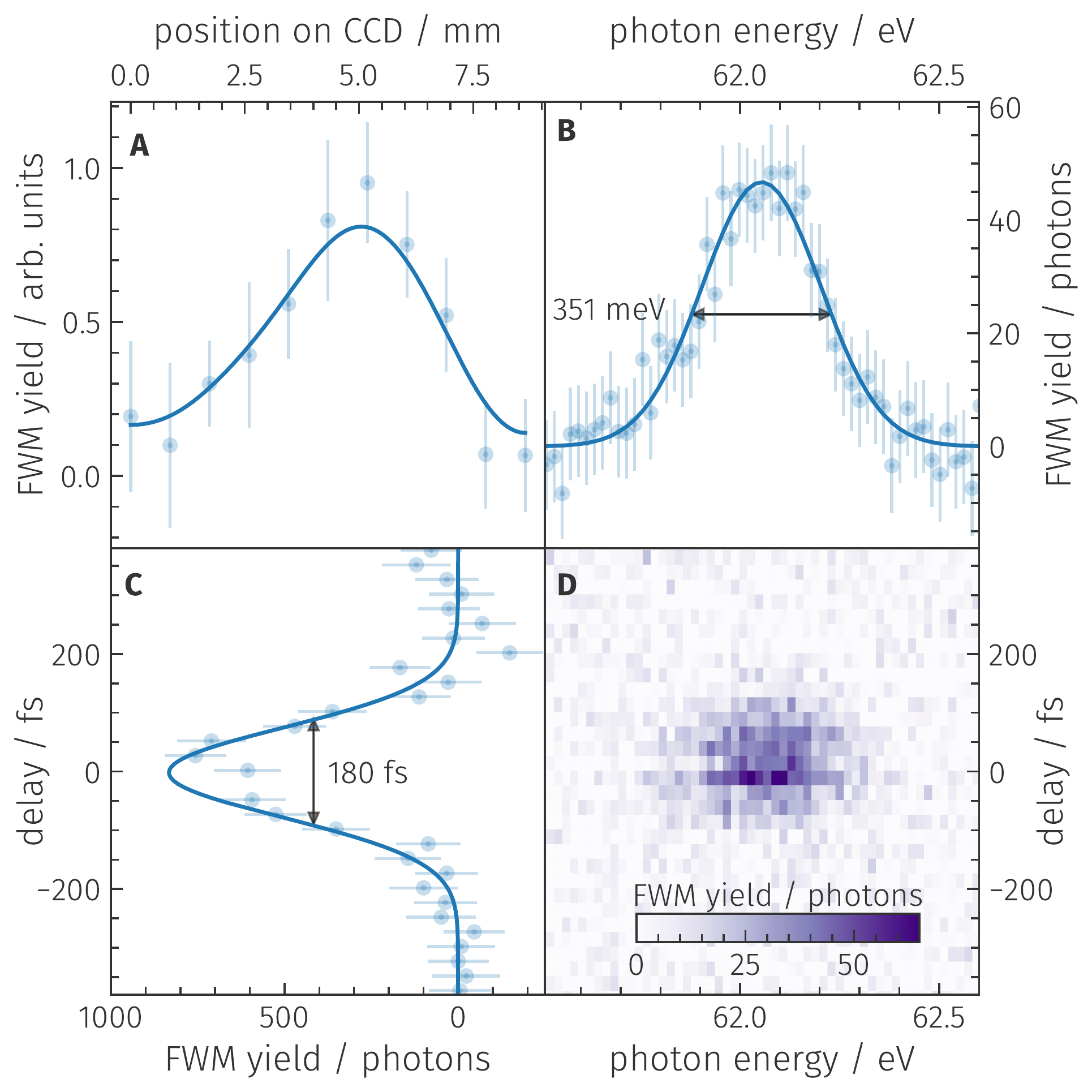}
\caption{\textbf{Characteristic details of the radiation generated in the nonlinear processes.} As a characteristic example we present details for the difference-frequency beam with the FEL tuned to three-photon resonance with the LiF core exciton at $\hbar\omega_\text{X} = 65.25$\,eV.
	\textbf{(A)} Sample profile of the generated difference-frequency beam on the CCD camera perpendicular to the dispersion direction of the grating. 
	The line shown is intended to guide the eye.
	\textbf{(B)} Spectral distribution of the difference-frequency photon yield with the FEL-NIR pulse delay set to zero, i.e. a cut through the density plot in panel \textbf{(D)} along a line at zero delay. \textbf{(C)} The total amount of generated difference-frequency photons plotted over the FEL-NIR pulse delay, i.e. a projection of the density plot in panel \textbf{(D)} on the delay axis. Blue curves in panels \textbf{(B)} and \textbf{(C)} represent Gaussian fits to the data and arrows indicate their FWHM. Error bars represent the standard deviation. \textbf{(D)} Density plot of the dependence of the spectral distribution of the difference-frequency photon yield on the FEL-NIR pulse delay.}
\label{fig3}
\label{fig4}
\end{figure}

Further corroboration of the FWM-nature of the signals shown in Fig.~\ref{fig2} comes from a detailed analysis of (i) the direction of propagation and divergence of the individual beams observed, (ii) their spectral and (iii) temporal behavior.
Details of these characteristics of the generated radiation beams are shown in Fig.~\ref{fig3}, exemplified for difference-frequency generation in three-photon resonance with the LiF $p$-type core exciton, obeying ($\omega_\text{X} - 2\omega_\text{I} = \omega_\text{exc}$). 

Regarding (i): In a FWM process, the emission direction of sum- and difference-frequency beams, which are emitted by the medium towards the vacuum side from where the driving radiation impinges (Fig.~\ref{setup}), are fixed by boundary conditions for the driving FEL and NIR fields at the crystal interface to vacuum, via the induced nonlinear polarization in the medium.
Fundamentally, this amounts to momentum conservation at the vacuum-medium interface, with the components of the driving fields' wave vectors along the interface on the vacuum side being the relevant quantities.
In the experiment the angle enclosed between the FEL and NIR beams is rather small ($\left|\theta_\text{X} - \theta_\text{I}\right| = 0.75^{\circ}$) and moreover $\omega_\text{X} \gg \omega_\text{I}$.
Hence the nonlinear reflection is expected closely along the reflected FEL beam with an estimated offset angle $\alpha$ of $\left|\alpha\right| < 0.04^{\circ}$ (see the supplement for more details). 
This angle is significantly smaller than the radial divergence of the FEL beam, which can be determined to be 0.12$^{\circ}$ using the characteristics of the focusing x-ray optics.
The experimental geometry thus requires a spectral separation of the FEL beam and the nonlinear reflection rather than a spatial separation (see Fig.~\ref{setup}).
An example for a difference-frequency beam profile on the CCD camera as projected on the direction perpendicular to the dispersion direction of the grating is shown in Fig.~\ref{fig3}(A).
The width of this beam amounts to $\approx 4\,$mm at half maximum, corresponding to a far field radial beam divergence of $\approx 0.09^{\circ}$.
As expected, the nonlinear emission cone is limited by the divergence of the focused XUV and NIR laser beams which drive the nonlinear processes.

Regarding (ii): In Fig.~\ref{fig4}(D) we show the spectral distribution of the frequency conversion yield over the relative delay of the FEL and NIR pulses.
A cut through this density plot along a line at zero delay, i.~e. the detailed spectral distribution of the $\omega_\text{X} - 2\omega_\text{I}$ difference-frequency photon yield, is displayed in Fig.~\ref{fig4}(B).
The distribution is centered at 62.05\,eV with a full width at half maximum (FWHM) of 351\,meV.
This is in close agreement with the expectation according to energy conservation requiring  $\hbar\omega = \hbar\omega_\text{X} - 2\hbar\omega_\text{I} = 62.15\,$eV.
The distribution closely matches the spectral width of the FEL radiation since the bandwidth of the NIR pulses is close to 100\,meV.

Regarding (iii): The dependence of the difference-frequency yield on the time delay between the FEL and NIR pulses in Fig.~\ref{fig4}(C) indicates that the nonlinear conversion process requires temporal overlap of the pulses.
This is expected for a nonlinear wave mixing process for which the induced nonlinear polarization decays on a time scale faster than the widths of the driving laser pulses.
The observed temporal profile represents the cross-correlation of the FEL and NIR pulses including the relative timing jitter of the pulses on the LiF sample.
A Gaussian fit to the yield indicates a FWHM of this cross-correlation of the pulses of $180\pm 22\,$fs.
Since the NIR laser pulse width was 21\,fs and the FEL pulse width was set to be less than 100\,fs, a substantial contribution of $\approx 140\,$fs to the width of the cross correlation curve is actually due to the relative arrival time jitter of the pulses on the LiF sample.

\paragraph*{Discussion}

Having firmly established that the signals we observe are due to FWM, we now turn our attention to the pronounced variation of the photon yield of the $\omega_\text{X} \pm 2\omega_\text{I}$ wave-mixing processes as a function of the driving FEL photon energy seen in Fig.~\ref{fig2}(C). How does  the nature of the resonances involved influence this yield?
Obviously, significant frequency conversion is present only in close one- or three-photon resonance with the $p$-type core exciton, giving rise to the characteristic two peak structure in both the $\omega_\text{X} + 2\omega_\text{I}$ and $\omega_\text{X} - 2\omega_\text{I}$ photon yields, separated by approximately $2\omega_\text{I}$ in Fig.~\ref{fig2}(C).
In contrast, one- or three-photon resonance exclusively with conduction band states does not give rise to detectable frequency conversion, i.e. no FWM yield was observed with an FEL photon energy setting higher than $\approx 64\,$eV and $\approx 67\,$eV for sum- and difference-frequency mixing, respectively.  

The influence of electronic correlation and in particular the formation of a core exciton next to the Li$^+$ $K$-edge absorption has been studied in the past~\cite{Haensel1968}, and first principles calculations clearly indicate that strong correlation effects between the core hole and the electrons are restricted to the first few eV of the XUV absorption starting at $\approx 60\,$eV~\cite{Kunz1984}. 
The nature of the electronic states involved in linear absorption and correspondingly in the linear reflectivity of LiF in Fig.~\ref{processes}(B) can explain what we observe in FWM.
In the initial state, the active electron is well localized on a Li atom in the crystal.
This also is true for the $p$-type core exciton, where the excited electron remains localized in the vicinity of that same Li atom~\cite{Olovsson2009a}.
This co-localization of the electron and hole can significantly enhance the third order nonlinear susceptibility $\chi^{(3)}$ of LiF, which is the quantity responsible for the the FWM processes we study.
In contrast, the higher energy conduction band electronic states can be described by delocalized Bloch states with the excited electron being able to move freely in the crystal.
The lack of co-localization in this case can in turn give rise to a strongly suppressed $\chi^{(3)}$.
This difference in electron-hole co-localization is reflected in the observed frequency conversion yield in Fig.~\ref{fig2}(C). 

Specifically, we note that we observe no significant difference-frequency generation when tuning the FEL photon energy to the 70.5\,eV feature in the linear reflectivity spectrum [Fig.~\ref{processes}(B)].
The corresponding feature in the LiF near $K$-edge absorption fine structure spectrum has been debated to be due to correlation effects, namely an electron polaron~\cite{Kunz1972,Sonntag1974}.
Furthermore, it has been speculated that this electron polaron is bound to the Li $1s$ core hole~\cite{Kunz1972}.
While this debate is about half a century old, ab-initio electronic structure calculations have to this day not been able to address this question.
Our experiment can address this question experimentally: In the hypothesized case of a bound polaron, an increase of the non-linear susceptibility and hence the FWM yield would be expected in one- and three-photon resonance with the 70.5 eV feature.
However, no significant FWM yield was observed either in the three-photon resonant sum frequency yield (Fig.~\ref{fig2}(C) blue curve around 67.4 eV) nor in the one-photon resonant difference frequency yield (Fig.~\ref{fig2}(C) orange curve around 70.5 eV).
This indicates the absence of co-localization of the active electron and core hole of the same order of magnitude as it is present for the core $p$-type exciton at 62.07\,eV.

Next, we strive to obtain a basic understanding of the pronounced sum- and difference-frequency conversion yield when in resonance with the $p$-type core exciton.
Given the strong localization and hence roughly atomic character of the core exciton, we model the third order nonlinear susceptibility tensor $\chi^{(3)}$, which is not known for LiF in the relevant photon energy range, by involving $\mathrm{Li}^+$ ionic states, namely the $(1s)^2$ ground state and excited states corresponding to the $p$-type  and to the $s$-type exciton.
This approach is in analogy to the early efforts to understand the linear absorption spectrum~\cite{Gudat1974}.
The details of the model are presented in the methods section and in the supplement.

\begin{figure}[h]
\centering
\includegraphics[width=0.75\textwidth]{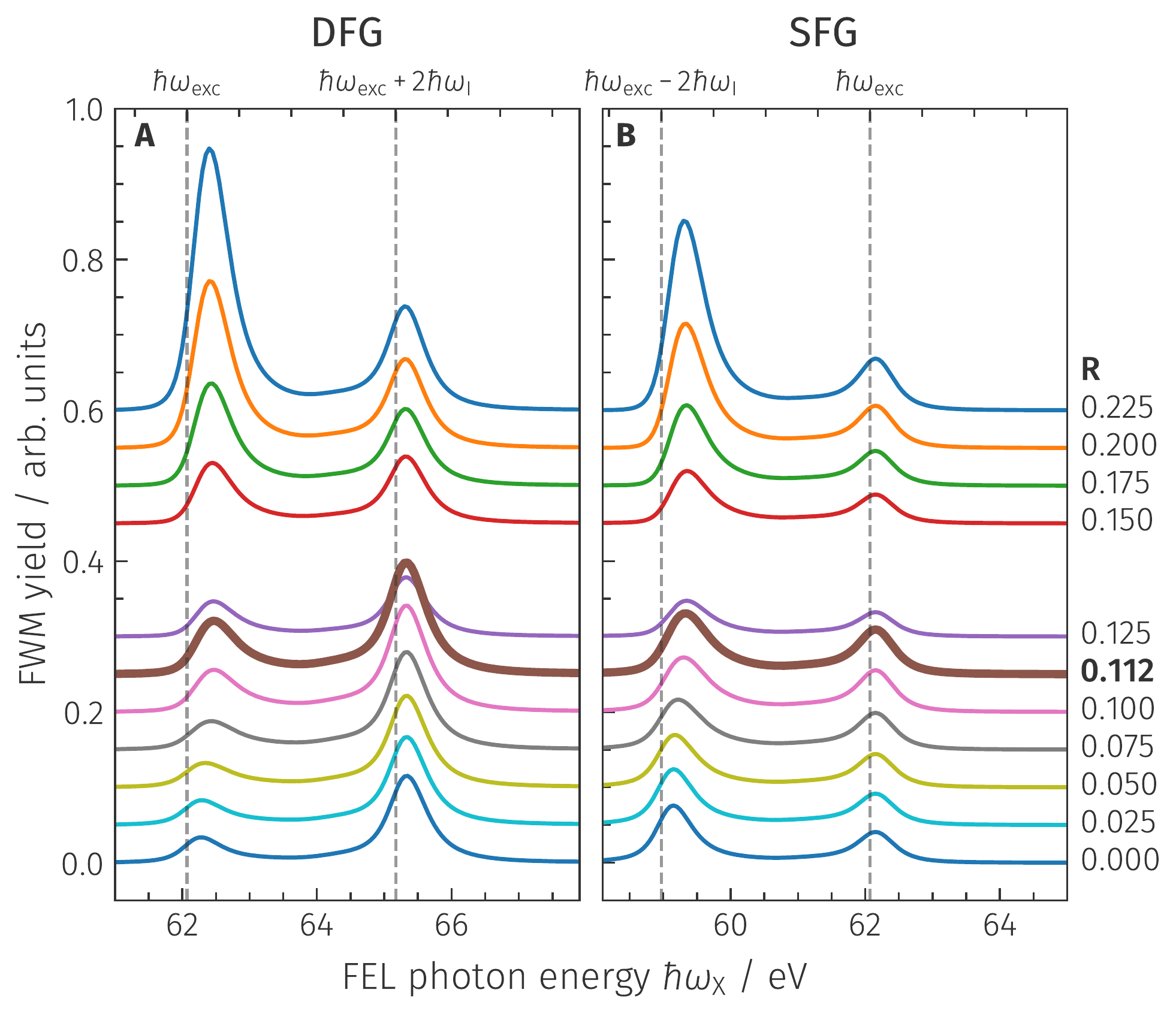}
\caption{\textbf{Calculated frequency conversion efficiencies.} Difference- \textbf{(A)} and sum-frequency mixing \textbf{(B)} yields calculated for the beams emitted from the LiF sample towards the vacuum side (reflective geometry, see Fig. \ref{geometry}). Shown are the dependencies of the efficiencies on the photon energy $\hbar\omega_\text{X}$ of the incident FEL radiation when scanned over the LiF $p$-type core exciton at $\hbar\omega_\text{exc} = 62.07$\,eV for a set of values of the  parameter $R$ starting at zero up to $R = 0.225$.
	$R$ quantifies the dipole coupling strength of the LiF $p$- to the $s$-type core exciton, the latter being separated in energy from the former by $1.1$~eV, i.~e. somewhat less than one NIR photon energy $\hbar\omega_\text{I} = 1.55$\,eV.
	For comparison purposes the conversion efficiency scales in panels \textbf{A} and \textbf{B} are chosen to be identical. Compared to the lower group of curves the calculated efficiencies for the upper group have been scaled by a factor of 0.5. In both panels we highlighted the efficiency curves for $R = 0.112$ which most closely reproduce the conversion yield structures found in the experiment [thick brown lines, see also Fig. \ref{fig2}(A)].}
\label{calc}
\end{figure}

Based on this $\chi^{(3)}$ model in conjunction with the experimental parameters, we obtain the conversion yields for the generated reflected sum- and difference-frequency beams as functions of the driving FEL photon energy as shown in Fig.~\ref{calc}.
It is assumed that the polarization of both the FEL and NIR radiation is in their common plane of incidence on the LiF crystal and their angle of incidence is 52$^\circ$ with respect to the crystal normal as it was in the experiment.
In Fig.~\ref{calc}(A) we present the dependence of the difference-frequency yield on the applied FEL photon energy.
Here, the ratio $R$ of the radial parts of the $\mathrm{Li^+}$ dipole matrix elements for the $(1s2s) - (1s2p)$ ($d_2$) and $(1s)^2 - (1s2p)$ ($d_1$) transitions $R = d_2/d_1$ serves as a parameter to characterize the relative influence of the states on $\chi^{(3)}$ and thus on the nonlinear polarization of the medium. The electric field strength of the incident XUV and NIR radiation are assumed as constant. In Fig.~\ref{calc}(B) we show the corresponding results for sum-frequency generation.

At $R = 0$ the LiF $s$-type core exciton does not influence the frequency conversion at all.
The resonance structure found thus exclusively originates from the driving laser radiation being either in one- ($\omega_\text{X} \approx \omega_\text{exc}$) or three-photon ($\omega_\text{X} \pm 2\omega_\text{I} \approx \omega_\text{exc}$) resonance with the $p$-type core exciton.
The slight offset of the frequency conversion maxima towards a higher FEL photon energy as compared with the $p$-type exciton position ($\omega_\text{exc} = 62.07\,$eV, Table~\ref{table} in the supplement and indicated by vertical lines in Fig.~\ref{calc}) can be attributed to the strong variation of the linear dielectric constant over the $p$-type exciton (Fig.~\ref{epsilon-calc} of the supplement) accompanied by strong absorption of either the driving or generated radiation on resonance.
For difference-frequency generation, a shifting of the conversion maximum at $\omega_\text{X} \approx \omega_\text{exc}$ towards a higher FEL photon energy is predicted with increasing $R$ (Fig.~\ref{calc}(A)), whereas the maximum at $\omega_\text{X} \approx \omega_\text{exc} + 2\omega_\text{I}$ does not shift.
Moreover, the conversion efficiency peak at $\omega_\text{X} \approx \omega_\text{exc}$ gains strongly in significance with rising $R$.
The underlying reason is a rising significance of two-photon resonance with the $s$-type exciton relative to one-XUV-photon resonance ($\omega_\text{X} \approx \omega_\text{exc}$) with the $p$-type exciton.
Two-photon resonance with the $s$-type exciton would occur at $\omega_\text{X} = 62.52\,$eV, compared to one-XUV-photon resonance with the $p$-type exciton at $\omega_\text{X} = 62.07\,$eV.
A rising value of R represents a stronger coupling between s- and p-type core excitations. Thus, the rising conversion efficiency and shift are a consequence of the s-type exciton providing an intermediate two-photon resonance in the three-photon DFG process, as the 1.55 eV NIR photons are not far off resonance with the energy difference of 1.1 eV between s- and p-type excitons.
In contrast, at three-photon resonance with the $p$-type exciton ($\omega_\text{X} \approx \omega_\text{exc} + 2\omega_\text{I}$) the frequency difference $\omega_\text{X} - \omega_\text{I}$ is well off the two-photon resonance with the $s$-type exciton.
This results in a negligible effect of two-photon resonance on the corresponding difference-frequency conversion yield and on the position of the yield maximum.

A similar interpretation holds for the changes observable in the sum-frequency conversion yield with rising values of the parameter $R$ in Fig.~\ref{calc}(B).
However, in this case the conversion yield maximum at $\omega_\text{X} \approx \omega_\text{exc} - 2\omega_\text{I}$, originating from three-photon resonance with the $p$-type exciton, is affected by two-photon resonance with the $s$-type exciton, whereas the $\omega_\text{X} \approx \omega_\text{exc}$ conversion maximum, representing one-XUV-photon resonance with the $p$-type exciton, is not.
Combined, the model particularly predicts the difference-frequency conversion yield maximum at $\omega_\text{X} \approx \omega_\text{exc}$ appearing at a slightly higher FEL photon energy than the corresponding sum-frequency yield maximum and links it to the NIR photons being able to cause a coupling between s- and p-type excitons for $R > 0$.

Comparing the outcome of our  model (Fig.~\ref{calc}) with the resonance structure observed in the experiment leads to an estimated $R$ of approximately $0.112$ when basing the comparison on the relative strengths of the frequency conversion yield maxima.
For the purpose of direct comparison with the experiment we show this calculated result in Fig.~\ref{fig2}(A) for both sum- and difference-frequency generation.
To illustrate the precision of this estimate: a transition moment ratio of $R = 0.225$ changes the resonance behavior of the frequency conversion processes completely in a way that is significantly different from the experimental result.
Similarly, $R = 0$ is not compatible with the experiment.
In the model $R \approx 0.112$ already results in a clear offset of the calculated difference-frequency conversion yield maximum at $\omega_\text{X} \approx \omega_\text{exc}$ to a higher FEL photon energy as compared to the corresponding sum-frequency yield maximum [Fig.~\ref{fig2}(A)].
Indeed, a hint at such a shift can also be found in the experimental data in Fig.~\ref{fig2}(C) at $\omega_\text{X} \approx \omega_\text{exc} \approx 62\,$eV.

Within the limitations of our simple $\chi^{(3)}$ model, we are able to understand the experimental observations quantitatively, in particular the energy positions and relative yields of the FWM processes in LiF when in resonance with Li$^+$ $1s$ core excitations, including the significance of a dipole coupling of the $p$- to the $s$-type excition in the nonlinear response. The model calculation reproduces the significant sum- and difference-frequency conversion yield when close to one- and three-photon resonance with the $p$-type core exciton.
While being in resonance with the $p$-type exiton is key for efficient frequency mixing, a two-photon resonance with the $s$-type exciton does affect the frequency conversion since a close to one-photon resonant, dipole allowed, coupling of the $s$- and $p$-type core excitons is present, mediated by the NIR laser pulses.
Thus, our findings further corroborate the assignment of the low energy shoulder of the linear reflection maximum observable at 61\,eV in Fig.~\ref{processes}(B) as due to an $s$-type exciton~\cite{Shirley2004,Olovsson2009a}. 

For NaCl, the observation of a complementary FWM process has recently been reported in a purely laser-based experiment, involving excitons related to the excited Na$^+$ $3s$ state with a hole in its closed $2p$ shell~\cite{Gaynor2021}.
This experiment revealed the presence of dark excited states which appear only in nonlinear spectroscopy besides bright excitonic excitations observable in linear absorption. 
A potential coherent lifetime of the nonlinear polarization induced in LiF as it was reported for NaCl on a time scale of 10\,fs in Ref.~\cite{Gaynor2021} is too short to be detectable with our present temporal resolution. 
With recent advances of pushing x-ray-optical relative pulse jitter at FELs to below 25\,fs~\cite{Sato2020} and the availability of both x-ray and optical pulses in the few fs or even as domain~\cite{Schulz2015,Duris2020}, we expect that such information for processes involving core excitations will also become accessible by our FEL-based approach, which can be immediately extended into the multi-keV photon energy range.

From an experimental point of view, we note that the third order nonlinear process studied in this work, i.~e. sum- and difference-frequency generation, allows for easy spectroscopic discrimination of the nonlinear signal from the incident XUV or x-ray radiation, which is particularly important when mixing harder x-rays with laser pulses in the optical spectral range due to the challenge posed by an increasingly small angular separation. Furthermore, we note that the observation of a nonlinear process in reflection from a solid sample-vacuum interface is convenient in particular in the XUV spectral range, since due to strong absorption only very thin samples can typically be used to detect the generated radiation propagating through the nonlinear medium - e.g. a sample of 50 nm thickness was used in Ref. \cite{Gaynor2021}.
We expect both these aspects as demonstrated here to be of relevance for the application of nonlinear x-ray spectroscopy in material science, where tailoring the samples under study to experiments is only viable to a limited extent.
Nevertheless, detecting transmitted radiation may be advantageous in selected cases where phase matching ($k_z - \tilde{k}_z = 0$ in equation~\ref{phasematching} in the supplement) can be exploited, provided that the propagation lengths of the FEL, optical and generated radiation in the sample can be made long enough prior to efficient absorption limiting the detectable signal. 
At interfaces of centrosymmetric crystals to vacuum or to other materials, a loss of inversion symmetry is commonly used to obtain surface or interface sensitivity via observation of second harmonic generation in the optical regime. As we have demonstrated, third order nonlinear signals can now be detected in one- and three-photon resonance with specific inner shell excitations.
Looking ahead, the combination of second and third order nonlinear frequency conversion involving x-rays will allow one to differentiate between bulk and surface specific contributions to the nonlinear polarization of a material.

Our work demonstrates that FWM under resonant conditions is a sensitive probe of charge localization, allowing for atomic specificity even when employing wavelengths significantly larger than the unit cell of the system of interest. While scaleable to hard x-rays, we note that even when carried out with soft x-rays, our approach 
can complement the atomic scale information on the valence charge distribution obtainable with wavelengths comparable to the interatomic spacing~\cite{Glover2012} and mesoscale information with a length scale defined by a transient grating period~\cite{Bencivenga2015,Bencivenga2019}.
Nonlinear spectroscopy with few-femtosecond x-rays gives access to otherwise forbidden, \emph{dark} transitions and hence information beyond the linear spectroscopy dominating current x-ray spectroscopy.
With the dynamics of transient localization and delocalization of charge at different atomic species being of fundamental importance not only for a multitude of processes in physics, chemistry and biology but also in materials design e.g. for light harvesting applications, we expect wave-mixing processes in resonance with inner shell excitations to become a particularly fruitful approach for future time-domain studies at FELs.

\paragraph*{Materials and Methods}
\subparagraph*{Experiment}
The experiment was carried out on the beamline FL24 of the FLASH\,2 free-electron laser using the MUSIX (MUlti-dimensional Spectroscopy and Inelastic X-ray scattering) endstation~\cite{Beye2019}.
A schematic view of the setup is shown in Fig.~\ref{setup}.
The FEL delivered pulses in bunch trains with 40 laser pulses per train at a train repetition rate of 10\,Hz.
The pulse separation within each train was 10\,\textmu sec, i.e. a pulse repetition rate of 100\,kHz.
The FEL beam was attenuated using a gas attenuator filled with $1.1\times 10^{-2}\,$mbar of Neon as well as a silicon attennuator foil of 411\,nm thickness.
The beam path was constrained by five beam diameter limiting irises, the first three with a diameter of 7.5\,mm, and the latter two with a diameter of 10.0\,mm and 10.5\,mm, respectively.
A focusing Kirkpatrick-Baez Active Optics System (KAOS) was used to focus the beam on the LiF sample~\cite{Raimondi2013}.
The spot size on the LiF crystal is estimated to be 150\,\textmu m FWHM in diameter as gauged with a Yttrium Aluminum Garnet (YAG) scintillation screen.
While the duration of the FEL pulses was not measured during the experiment it was tuned to be less than 100\,fs.
The spectral bandwidth of the FEL pulses amounted to $\approx 0.6 \%$ of the photon energy, i.e. $\approx 0.4\,$eV in the photon energy range used in the experiment.
The angle of incidence of the beam on the LiF sample with respect to its surface normal was 52$^\circ$.
The energy of individual FEL pulses on the sample was $130 \pm 25\,$nJ [see also Fig.~\ref{fig2}(D)].
This energy and the corresponding number of photons impinging on the LiF sample per pulse were determined by using an x-ray gas monitor detector upstream of the FEL beam and multiplying these data by an estimated beamline transmission up to the sample~\cite{Sorokin2019}.
The supplement provides further details on the transmission estimates of the beamline. 

The optical laser system was based on optical parametric chirped pulse amplification (OPCPA) synchronized to the FEL pulses and operated at a fixed photon energy of 1.55\,eV.
It delivered pulse trains identical in structure to those of the FEL with pulses of 21\,fs in width (FWHM).
The NIR laser beam was focused on the LiF sample using a lens pair with -250\,mm and 750\,mm focal lengths, respectively.
After passing through a $\lambda/2$ wave plate and polarizer for adjusting the pulse energy and a second half-wave plate for turning its polarization, the linearly polarized NIR beam was aligned collinear with the FEL beam path by reflecting it off a flat 90$^\circ$ silver coated turning mirror which passed the FEL beam through a center hole (diameter: 5\,mm).
Thereby the NIR laser beam enclosed a small angle of $0.75 \pm 0.03^\circ$ with the FEL beam, propagating in the plane of incidence spanned by the FEL beam and the LiF surface normal, which coincided with one of the LiF crystal axes.
This plane of incidence is assumed to be spanned by the $x,z$-axes of a suitably chosen coordinate system as shown in Fig.~\ref{geometry}.
The two LiF crystal axes in the interface plane to vacuum, i.~e. the two red axes in Fig.~\ref{geometry}, enclosed an angle of $(22.5 \pm 2.5)^\circ$ with the $x,y$-axes of this coordinate system, respectively.
The polarization vectors of the FEL and NIR pulses were located in the plane of incidence.
The NIR laser pulse energy on the LiF target was $(68.0 \pm 1.5)\,$\textmu J with the spots of the FEL and NIR beams spatially overlapping.

The LiF sample was a single, (100) cut crystal with the crystal surface polished.
With a thickness of 3\,mm it was transparent to the NIR laser beam, whereas the FEL beam was practically absorbed within $\approx 100\,$nm into the crystal, thus limiting the sum- and difference-frequency generation to a narrow stretch below the crystal surface.
The NIR, FEL and generated sum- and difference frequency beams reflected off the LiF sample were dispersed by a reflection grating and directed to a CCD camera positioned behind an aluminum filter (thickness: 200\,nm) which blocked the NIR beam and stray light, but also inhibited the detection of FWM signals beyod the aluminium $K$-edge at 72.7\,eV.
The specular reflection of the FEL beam was blocked by a narrow beam stop placed directly in front of the CCD camera (see Fig.~\ref{setup}).

The dispersion grating with spherical shape and varied line spacing was used to separate the FEL radiation from the generated sum- and difference-frequency radiation~\cite{Beye2019}.
It was operated at grazing incidence (grazing angle: 2.1$^\circ$) and imaged the radiation spectrum onto the CCD camera.
Due to the large radius of curvature of the grating ($\approx 13.9\,$m) it had little influence on the divergence of the sum- and difference-frequency beams on their way from the LiF sample to the camera along the direction orthogonal to the dispersion direction of the grating.
The length of the spectrometer setup, i.~e. the separation of the LiF sample from the camera via the grating, was 1.33\,m.

The back-illuminated CCD camera (Greateyes Model GE-VAC 2048 2048) used to detect the XUV radiation was mounted in the vacuum chamber~\cite{Beye2019}.
Its detector consisted of 2048 by 2048 square pixel of 13.5\,\textmu m in size~\cite{Beye2019}. 
For the experiment, four detector pixel were binned along the direction of dispersion of the spectrometer grating and 64 along the orthogonal direction to optimize the camera sensitivity.
To speed up the readout rate, only 11 binned pixel perpendicular to the dispersion direction of the grating were read out.
Thus, each stored camera image consisted of $512 \times 11$ image points.
The detector accumulated the light of individual pulse trains, i.e. of 40 laser pulses before being read out.
Limited by the detector read-out speed, a frame rate of 5\,Hz was achieved, meaning the sum- and difference frequency light generated by every second FEL pulse train was captured.

In the experimental runs, the FEL photon energy was tuned in steps in the range between approximately 58\,eV and 72\,eV. 
The FWM yields shown in Fig. \ref{fig2}(B,~C) were measured over the course of about 100 hours with individual data-sets recorded for at most 30 to 60 minutes on a single spot on the LiF sample at each FEL photon energy setting.
At each energy setting, the delay between the XUV and NIR pulses was scanned within a 1-2\,ps interval around the time-overlap.
This allowed to account for slow drifting of the XUV pulses' arrival time relative to that of the NIR pulses, which would have affected the FWM efficiency.
The CCD camera recorded the radiation generated by FWM spectrally dispersed by the spectrometer grating. 
The recorded spectra were scaled to estimate the absolute number of photons actually generated by FWM in the LiF sample (see the supplement for details).
Fluctuations in optical stray light, generating a varying offset to each spectrum, were compensated by subtracting the average signal recorded in the FEL beam block shadow from each spectrum.
Intensity fluctuations in the FEL beam were mitigated by normalizing the spectra such that the FEL-stray light level of each spectrum equals the ensemble average.
From these spectral distributions a mean straylight spectrum was determined by averaging over all spectra recorded at FEL-NIR pulse delays larger than 250\,fs, leaving only the FWM signal together with a small, however fluctuating, residual background level.
These spectra where then sorted by the XUV-NIR-delay into delay-time bins of 25\,fs with an average FWM-spectrum computed for each bin.
The result is a two-dimensional map of the FWM signal over the XUV-NIR-delay and the energy distribution of the spectrally dispersed generated photons, as shown in Fig.~\ref{fig4}(D).
While the average spectrum was calculated by the simple mean for each photon energy, an estimate of the uncertainty of this mean was computed as the standard deviation of the averaged photon yields divided by the square root of the number of spectra in the respective bin [see Fig.~\ref{fig4}(B)].

The linear reflection of LiF shown in Fig.~\ref{processes}(B) was measured separately at the PM3 beamline of the BESSY II facility~\cite{Kachel2016} using tunable synchrotron radiation linearly polarized in the plane of incidence of the beam.
The angle of incidence on the LiF single crystal was 50$^\circ$ with respect to the surface normal of the sample.

\subparagraph*{Theoretical modeling of 3\textsuperscript{rd} order FWM}

The nonlinear reflection off a medium with a plane boundary to vacuum in the $x,y$-plane of a suitably chosen coordinate system (see Fig.~\ref{geometry}) is connected with the generated electric field on the vacuum side. In the monochromatic, plane wave limit this electric field can be represented by: 

\begin{equation}\label{reffield}
	\mathbf{E}_r = \mathbf{A}_r \exp {\left(i \mathbf{k}_r \mathbf{x} - i \omega t \right)}
\end{equation}

with $\mathbf{A}_r$ its amplitude and $\mathbf{k}_r = (k_x, 0, k_{rz})$ the wave vector on the vacuum side ($k_{rz} > 0$) satisfying the dispersion relation $\mathbf{k}_r^2 = k_x^2 + k_{rz}^2 = \left( \omega / c \right)^2$ ($c$: the speed of light). The amplitude $\mathbf{A}_r$ can be found to be (for the derivation see the supplement)

\begin{equation}
	\label{refamp}
	\mathbf{A}_r = \frac{4\pi}{\tilde{k}_z + k_z} \left(\frac{\left(\tilde{k}_z P_x + k_x P_z \right)}{\tilde{k}_z - \epsilon k_{rz}}
	\begin{bmatrix}
		k_{rz} \\
		0 \\
		-k_x
	\end{bmatrix}
	- \left(\frac{\omega}{c}\right)^2 \frac{P_y}{\tilde{k}_z - k_{rz}}
	\begin{bmatrix}
		0 \\
		1 \\
		0
	\end{bmatrix}\right).
\end{equation}

This expression presumes a monochromatic, plane nonlinear polarization wave $\mathbf{P} \exp{\left(i \mathbf{kx} - i \omega t \right)}$ propagating into the medium which is assumed to be located in the half space $z < 0$ with frequency $\omega$ and wave vector $\mathbf{k} = (k_x, 0, k_z)$ chosen to be located in the $x,z$-plane (see Fig.~\ref{geometry}). The real and the imaginary part of $k_z$ are negative numbers, representing propagation of the nonlinear polarization into the medium and absorption of the fields driving it. $\mathrm{\tilde{\mathbf{k}}} = (k_x, 0, \tilde{k}_z)$ in equation~\ref{refamp} is the wave vector of the generated electric field at frequency $\omega$ propagating in the nonlinear medium. It satisfies the dispersion relation $\tilde{\mathbf{k}}^2 = k_x^2 + \tilde{k}_z^2 = \left( \omega / c \right)^2 \epsilon\!\left(\omega\right)$, where $\epsilon\!\left(\omega\right)$ is the dielectric constant of the medium at the frequency $\omega$ of the generated field. In the medium it propagates with the nonlinear polarization and experiences linear absorption, meaning the real and the imaginary parts of $\tilde{k}_z$ are negative numbers. $k_x$, the wave number component in the plane of the surface of the nonlinear medium, which appears in the expressions for $\mathbf{k}_r$, $\mathbf{k}$ and $\tilde{\mathbf{k}}$, is a real number and is fixed by boundary conditions for the FEL and NIR fields, which drive the nonlinear polarization, at the interface of the medium with vacuum.

The components of the wave-vector $\mathbf{k}$ of the nonlinear polarization wave and its frequency $\omega$ are fixed by the frequencies and wave-vectors of the fields driving the nonlinear polarization and by the specific nonlinear process, i.e. in our case third order sum- and difference-frequency mixing, respectively. Provided $\omega_\text{X}$, $\mathbf{k}_\text{X}$ and $\omega_\text{I}$, $\mathbf{k}_\text{I}$ are the frequencies and wave vectors of the driving FEL and NIR laser fields in the medium, respectively, $\omega$, $\mathbf{k}$ are defined by
\begin{align*}
	\omega &= \omega_\text{X} \pm 2\omega_\text{I} \\
	\mathbf{k} &= \mathbf{k}_\text{X} \pm 2\mathbf{k}_\text{I}
\end{align*}
This assumes one FEL photon and two NIR photons are involved in the mixing process as was the case in the experiment.

The amplitude $\mathbf{P}$ of the nonlinear polarization wave in equation \ref{refamp} is set by the third order nonlinear susceptibility tensor $\mathbf{\chi^{(3)}}$ of LiF governing the experimentally studied sum- and difference frequency generation. For the cubic LiF crystal structure, only few of the cartesian components of $\mathbf{\chi}^{(3)}$ are different from zero and independent of each other \cite{Butcher1990}. In a coordinate system aligned with the crystal axes, a set of independent non-zero components is $\chi^{(3)}_{\mathrm{xxxx}}, \chi^{(3)}_{\mathrm{xyyx}}, \chi^{(3)}_{\mathrm{xyxy}}, \chi^{(3)}_{\mathrm{xxyy}}$. We use atomic transition matrix elements among $\mathrm{Li}^+$ states for the purpose of a basic approximation of $\mathbf{\chi^{(3)}}$. Therefore, full rotational symmetry prevails for the tensor elements, not just cubic symmetry. This adds an additional constraint to the tensor components, namely $\chi^{(3)}_{\mathrm{xxxx}} = \chi^{(3)}_{\mathrm{xyyx}} + \chi^{(3)}_{\mathrm{xyxy}} + \chi^{(3)}_{\mathrm{xxyy}}$ \cite{Butcher1990}. This extended model symmetry and the fact that the LiF dielectric constant is a scalar quantity also means the Cartesian coordinate system tied to the experimental plane of incidence in Fig. \ref{geometry} is equivalent to the coordinate system spanned by the cubic crystal axes. Then, independent of the direction of polarization of the NIR laser beam, the nonlinear polarization component $P_y$ in Eq.~\ref{refamp} is equal to zero,  meaning that the field generated is always polarized in the plane of incidence.

The $p$-type core exciton of LiF at 62.07\,eV observed in linear absorption spectra has been attributed to derive from Li$^+$ dipole allowed transitions from the $(1s)^2$ ground state to excited states $(1s2p,\, m_l = 0,\, \pm 1)$. A second, much smaller resonance structure on the low energy side of this $p$-type resonance has been tentatively associated with a $(1s)^2 - (1s2s)$ Li$^+$ transition ($s$-type exciton), not dipole allowed in the free ion \cite{Fields1977,Olovsson2009a}. A potentially sizable dipole allowed transition matrix element between the $(1s2p,\, m_l = 0,\, \pm 1)$ and $(1s2s)$ states may, however, influence the third order nonlinear susceptibility $\chi^{(3)}$ substantially. We thus model $\chi^{(3)}$ of the LiF crystal using just these Li$^+$ ionic states, ignoring any potential contribution of conduction band states. Our simple model is intended to qualitatively represent the experimentally observed dependence of the reflected sum- and differences frequency yields on the FEL photon energy [Fig.~\ref{fig2}(C)]. Thus, only the ratio of the dipole matrix elements for the Li$^+$ transitions $(1s2s) - (1s2p,\, m_l = 0,\, \pm 1)$ and $(1s)^2 - (1s2p,\, m_l = 0,\, \pm 1)$ will be relevant for the characterization of $\chi^{(3)}$. Moreover, we will retain the full atomic rotational symmetry for the dipole matrix elements, ignoring crystal field effects. This reduces the number of adjustable parameters in an expression for $\chi^{(3)}$ to the ratio $R$ of the radial parts of these dipole matrix elements. The excitation energies of the $p$- and $s$-type core excitons and their widths, assumed to be homogeneous widths, are extracted from the linear reflection data in Fig.~\ref{processes}(B). The data used to determine the components of the third order nonlinear susceptibility are summarized in the supplemental material and the expression used for our model nonlinear susceptibility is taken from Ref.~\cite{Butcher1990} (page 93).

A  more sophisticated theoretical treatment of the third order nonlinear susceptibility tensor of LiF in this photon energy range is beyond the scope of this work.
It will have to consider a potential influence of  conduction band states, account for the excitonic structure more precisely, and take into account the lower cubic crystal symmetry which we replaced by full rotational symmetry in our model.
We note that the lower symmetry may have an influence on the relative strengths of the frequency conversion yields, since in that case the angle enclosed by the plane of incidence of the laser beams with the LiF crystal axes in the surface plane of the crystal matters (see the supplement).
Also, we have used a monochromatic approach whereas the laser pulses have a short pulse duration and possibly a substantial phase modulation accompanied by a corresponding spectral bandwidth which is of the order of the widths of the relevant LiF resonance structures.


\nocite{Milgram1962,Jackson1998,Braune2018,Henke,Henke1993}


\clearpage

\paragraph*{Acknowledgements:} 
%
We acknowledge DESY (Hamburg, Germany), a member of the Helmholtz Association HGF, for the provision of experimental facilities. Parts of this research were carried out at FLASH, beamline FL24 using the MUSIX end station. Beamtime was allocated for proposal F-20181181.
Other parts of the research were carried out at the PM3 beamline at the BESSY~II electron storage ring operated by the Helmholtz-Zentrum Berlin für Materialien und Energie.
	
\noindent \textbf{Funding:}

Helmholtz foundation grant VH-NG-1105 (R.Y.E, J.S., P.S.M., M.B.)
\newline
\textbf{Author contributions:}

Conceptualization: H.R., S.E., M.B.
 
Experiment: H.R., R.Y.E., J.O.S., P.S.D., M.C.B., M.K., N.E., S.D., G.B., U.E., D.S., C.v.K.S.

Data analysis: R.Y.E., H.R

Theoretical model \& calculations: H.R.

Interpretation: H.R., S.E., R.Y.E., M.B.

Writing -- original draft: H.R., D.S., S.E. (with input from all authors)
\newline
\textbf{Competing interests:} Authors declare that they have no competing interests.
\newline
\textbf{Data and materials availability:} \url{10.5281/zenodo.5723994}
\newline


\clearpage

\noindent \textbf{Figures}

\clearpage

\clearpage

\clearpage

\clearpage

\clearpage

\appendix

\section*{Supplementary Materials}

\renewcommand{\thefigure}{S\arabic{figure}}
\setcounter{figure}{0}

\renewcommand{\theequation}{S\arabic{equation}}
\setcounter{equation}{0}

\noindent \textbf{PDF file includes:}

Supplementary Text

Figs. S1 to S2

Table S1

References (1 to 7)

\clearpage

\paragraph*{Frequency conversion: monochromatic plane wave limit}

Throughout the following derivation we will employ the cgs system of units. The Maxwell equations in the frequency domain used to determine the propagation of the nonlinearly generated electromagnetic waves in the LiF crystal while applying intense driving radiation fields are:
\begin{linenomath*}
	\begin{align}
		\nabla\, \mathsf{x}\, \mathbf{E}_\omega\!\left(\mathbf{x}\right)& = i \frac{\omega}{c} \mathbf{B}_\omega\!\left(\mathbf{x}\right)& \nabla \mathbf{B}_\omega\!\left(\mathbf{x}\right) = 0 \label{a}\\
		\nabla\, \mathsf{x}\, \mathbf{B}_\omega\!\left(\mathbf{x}\right)& = -i \frac{\omega}{c} \mathbf{D}_\omega\!\left(\mathbf{x}\right)& \nabla \mathbf{D}_\omega\!\left(\mathbf{x}\right) = 0 \label{b}.
	\end{align}
\end{linenomath*}
$\omega$ represents the frequency of the generated electromagnetic wave, $\mathbf{E}_\omega\!\left(\mathbf{x}\right)$ its electric field, $\mathbf{B}_\omega\!\left(\mathbf{x}\right)$ the corresponding magnetic induction and $\mathbf{D}_\omega\!\left(\mathbf{x}\right)$ the dielectric displacement. $c$ is the speed of light. The equations above assume the magnetic permeability of the nonlinear medium to be equal to $1$. The dielectric displacement is linked to the electric field strength via:
\begin{linenomath*}
	\begin{equation}
		\mathbf{D}_\omega\!\left(\mathbf{x}\right) = \epsilon\!\left(\omega\right) \mathbf{E}_\omega\!\left(\mathbf{x}\right) + 4\pi \mathbf{P}_\omega\!\left(\mathbf{x}\right)
		\label{dielectric}
	\end{equation}
\end{linenomath*}
with $\epsilon\!\left(\omega\right)$ the linear dielectric constant of the medium at the frequency $\omega$ of the generated electromagnetic wave and $\mathbf{P}_\omega\!\left(\mathbf{x}\right)$ the nonlinear contribution to the polarization of the medium at that frequency. The polaritation wave, in our case, is driven by the propagation of the FEL and NIR laser fields in the LiF crystal. We assume $\epsilon\!\left(\omega\right)$ to be a scalar quantity since the LiF crystal is of cubic symmetry. Absorption of LiF at the frequency of the generated wave renders $\epsilon\!\left(\omega\right)$ a complex quantity with its imaginary part being positive. On the vacuum side from where the FEL and NIR laser beams enter the crystal the same Maxwell equations above apply with $\epsilon\!\left(\omega\right) = 1$ and $\mathbf{P}_\omega\!\left(\mathbf{x}\right) = 0$.

The Maxwell equations together with relation \ref{dielectric} for the dielectric displacement can be combined into an inhomogeneous wave equation for the electric field $\mathbf{E}_\omega\!\left(\mathbf{x}\right)$:
\begin{linenomath*}
	\begin{equation}
		\Delta \mathbf{E}_\omega + \epsilon\!\left(\omega\right) \left(\frac{\omega}{c}\right)^2 \mathbf{E}_\omega = -4\pi \left[ \left(\frac{\omega}{c}\right)^2 \mathbf{P} - \frac{1}{\epsilon\!\left(\omega\right)} \mathbf{k}\left(\mathbf{k} \mathbf{P} \right) \right] \exp{\left(i \mathbf{kx}\right)},
		\label{waveeq}
	\end{equation}
\end{linenomath*}
with solutions subject to the additional restriction:
\begin{linenomath*}
	\begin{equation}
		\nabla \mathbf{E}_\omega = - \frac{4\pi i}{\epsilon\!\left(\omega\right)} \mathbf{k}  \mathbf{P}\exp{\left(i \mathbf{kx}\right)}.
		\label{divergence}
	\end{equation}
\end{linenomath*}
Relations \ref{waveeq} and \ref{divergence} assume $\mathbf{P}_\omega\!\left(\mathbf{x}\right)$ to be a plane wave $ \mathbf{P}_\omega\!\left(\mathbf{x}\right) = \mathbf{P} \exp \left(i\mathbf{kx}\right)$, implying the driving FEL and NIR laser fields are plane waves. The wave vector $\mathbf{k}$ of the polarization wave is fixed by the specific nonlinear conversion process and by the wave vectors of the driving laser fields in the medium. We further assume the polarization wave to propagate in the $(x,z)$-plane of a suitably chosen coordinate system with $\mathbf{k} = \left(k_x, 0, k_z\right)$ determined by the common plane of incidence of the FEL and NIR radiation on the LiF crystal. $\mathbf{P}$ represents the constant amplitude of the polarization wave. We do not account for any potential nonlinear polarization of the medium bound to the vacuum-material interface.

The Maxwell equations \ref{a} and \ref{b} imply specific boundary conditions at the vacuum-medium interface (the $x,y$-plane). Namely, the magnetic induction has to be continuous across the boundary. Also the $x$- and $y$-components of the electric field (i.e. the components parallel to the interface) and the $z$-component of the dielectric displacement (i.e. the component perpendicular to the interface) have to be continuous across the interface. In the medium the generated electromagnetic wave propagates with the polarization wave $\mathbf{P}_\omega\!\left(\mathbf{x}\right)$ into the medium, i.e. towards $z = -\infty$. The generated wave on the vacuum side propagates towards $z = +\infty$. There is no wave on the vacuum side propagating towards the interface at $z = 0$ at the frequency $\omega$. Since the wave vector of the polarization wave $\mathbf{k} = \left(k_x, 0, k_z\right)$ has zero $y$-component so the wave vectors of the generated waves in vacuum and in the medium have a zero $y$-component. The electric field of the generated wave on the vacuum side can thus be represented by  $\mathbf{E}_{r}\!\left(\mathbf{x}\right) = \mathbf{A}_r \exp\left(i \mathbf{k}_r\mathbf{x}\right)$ with $\mathbf{k}_r = \left(k_{rx}, 0, k_{rz}\right)$, transverse electric field amplitude $\mathbf{A}_r$ ($\mathbf{k}_r \mathbf{A}_r = 0$) and $k_{rz} > 0$. The dispersion relation $\mathbf{k}_r^2 = \left(\omega / c\right)^2$ ties the $x$- and $z$-components of the wave vector together.

The inhomogeneous wave equation \ref{waveeq} in the medium for this geometry is solved with the Ansatz
\begin{linenomath*}
	\begin{equation}
		\mathbf{E}_M\!\left(\mathbf{x}\right) = \tilde{\mathbf{E}}\left(z\right) \exp{(i \tilde{k}_x x)}.
	\end{equation}
\end{linenomath*}
It results in an inhomogeneous ordinary second order differential equation for the amplitude $\tilde{\mathbf{E}}\left(z\right)$
\begin{linenomath*}
	\begin{equation}
		\frac{\mathrm{d}^2 \tilde{\mathbf{E}}\left(z\right)}{\mathrm{d}z^2} +  \tilde{k}_z^2 \tilde{\mathbf{E}}\left(z\right) = \frac{4\pi}{\epsilon\!\left(\omega\right)} \left[ \mathbf{k}\left(\mathbf{k} \mathbf{P} \right) - \epsilon\!\left(\omega\right) \left( \frac{\omega}{c}\right)^2 \mathbf{P} \right] \exp{\left(i k_z z\right)}.
		\label{ordinaryweq}
	\end{equation}
\end{linenomath*}
$\tilde{k}_z$ used in Eqn.~\ref{ordinaryweq} is set via the dispersion relation $\tilde{k}_z^2 =  \epsilon\!\left(\omega\right) \left(\omega / c\right)^2 - \tilde{k}_x^2$ with the real and imaginary parts of $\tilde{k}_z$ chosen to be negative in order to make sure the generated electric field in the nonlinear medium propagates with the polarization wave towards $z = - \infty$ and force absorption in the medium. With these restrictions the general solution of Eq.~\ref{ordinaryweq} reads
\begin{linenomath*}
	\begin{equation}
		\tilde{\mathbf{E}}\left( z \right) = \tilde{\mathbf{A}} \exp{(i \tilde{k}_z z)} + \frac{\mathbf{H}}{\tilde{k}_z^2 - k_z^2} \exp{\left(i k_z z\right)},
		\label{mediumresult}
	\end{equation}
\end{linenomath*}
with
\begin{linenomath*}
	\begin{equation}
		\mathbf{H} = \frac{4\pi}{\epsilon\!\left(\omega\right)} \left[ \mathbf{k}\left(\mathbf{k} \mathbf{P} \right) - \epsilon\!\left(\omega\right) \left( \frac{\omega}{c}\right)^2 \mathbf{P} \right].
	\end{equation}
\end{linenomath*}
The amplitude vectors $\mathbf{A}_r$ of the electric field on the vacuum side (Eq.~\ref{reffield} of the main text) and $\tilde{\mathbf{A}}$ are still free constants. They are fixed by the boundary conditions at the vacuum-medium interface, by the transversality of the electric field in vacuum and by Eq.~\ref{divergence} in the medium. The boundary conditions can only be satisfied provided the still free wave vector components $\tilde{k}_x$ and $k_{rx}$ satisfy the condition $\tilde{k}_x = k_{rx} = k_x$, i.e. they have to be equal to the $x$-component of the wave vector of the nonlinear polarization wave. Via the dispersion relations on the vacuum side and in the medium then also the $z$-components $\tilde{k}_z$ and $k_{rz}$ are fixed subject to the constraints $k_{rz} > 0$ and the real and imaginary parts of $\tilde{k}_z$ have to be negative.

Involving these constraints results in the explicit form for the electric field amplitude $\mathbf{A}_r$ on the vacuum side given in Eq.~\ref{refamp} of the main text and in
\begin{linenomath*}
	\begin{align}
		\tilde{\mathbf{A}} =& \frac{4\pi}{\epsilon\!\left(\omega\right) \left[ \tilde{k}_z - \epsilon\!\left(\omega\right) k_{rz} \right]} \left\{ P_x + \left( k_z - \epsilon\!\left(\omega\right) k_{rz} \right) \frac{[\mathbf{k}\, \mathsf{x}\, \mathbf{P}]_y}{\tilde{k}_z^2 - k_z^2}  \right\}
		\begin{bmatrix}
			\tilde{k}_z \\
			0 \\
			-k_x
		\end{bmatrix}
		+  \nonumber \\
		&4\pi \left(\frac{\omega}{c}\right)^2 \frac{k_z - k_{rz}}{\tilde{k}_z - k_{rz}}\: \frac{P_y}{\tilde{k}_z^2 - k_z^2}
		\begin{bmatrix}
			0 \\
			1 \\
			0
		\end{bmatrix}
	\end{align}
\end{linenomath*}
The result Eq.~\ref{mediumresult} for the dependence of the electric field in the medium on the $z$-coordinate may be rewritten in the form
\begin{linenomath*}
	\begin{equation}
		\tilde{\mathbf{E}}\left( z \right) = \left[\tilde{\mathbf{A}} + \frac{\mathbf{H}}{\tilde{k}_z^2 - k_z^2} + \mathbf{H} \frac{\exp{\![i (k_z - \tilde{k}_z) z]} - 1}{\tilde{k}_z^2 - k_z^2} \right] \exp{(i \tilde{k}_z z)}
		\label{phasematching}
	\end{equation}
\end{linenomath*}
to emphasize the role of phase matching in the nonlinear process provided the wave vectors $\mathbf{k}$ of the nonlinear polarization wave in the medium and $\mathbf{\tilde{k}}$ of the generated wave become equal (meaning $\tilde{k}_z = k_z$ in Eq.~\ref{phasematching}). In the limit of phase matching the $z$ independent term in square brackets in Eq.~\ref{phasematching}, namely $\tilde{\mathbf{A}} + \mathbf{H} / (\tilde{k}_z^2 - k_z^2 )$, does not become singular. 
This can be seen when explicitly evaluating it using the expressions for $\mathbf{H}$ and $\tilde{\mathbf{A}}$ given above. Phase matching means the $z$ dependent term in square brackets in Eq.~\ref{phasematching}, i.e. the amplitude of the electric field of the generated wave in the medium grows in proportion to the propagation length $z$ in the medium. However, one has to keep in mind that absorption in the medium will counteract the buildup of the electric field. Absorption enters via the imaginary part of $\tilde{k}_z$ in $\exp{(i \tilde{k}_z z)}$ in the expression for the electric field in Eq.~\ref{phasematching}. After a certain propagation length absorption will always win over the linear buildup in $z$ of the wave's amplitude.

\paragraph*{The directions of emission of the sum- and difference-frequency beams on the vacuum side}

In the plane wave approximation used here the wave vector components of the XUV and NIR beams impinging on the LiF crystal in the surface plane of the crystal determine the corresponding component of the wave vector of the polarization wave in the crystal.
We represent these wave vectors by $\mathbf{k}^{v\text{X}} = (k^{v\text{X}}_x, 0, k^{v\text{X}}_z)$ and $\mathbf{k}^{v\text{I}} = (k^{v\text{I}}_x, 0, k^{v\text{I}}_z)$ for the XUV and NIR beams, respectively, using the coordinate system defined in Fig.~\ref{geometry}.
Boundary conditions for the electromagnetic fields imply that the $x$-components of the wave vectors of the XUV and NIR beams (in plane components), which drive the nonlinear processes, do not change when passing into the LiF crystal. Then the $x$-components of the wave vectors of the polarization waves in the medium are $k_x = k^{v\text{X}}_x \pm 2 k^{v\text{I}}_x$, respectively.
According to the previous section $k_x$, in turn, equals the $x$-component of the wave vectors of the generated electromagnetic fields on the vacuum side ($k_{rx} = k_x$).
Based on the dispersion relation for the sum- and difference-frequency fields on the vacuum side they thus propagate along the wave vector $\mathbf{k}_r = \left(k_x, 0, \sqrt{(\omega / c)^2 - k_x^2}\right)$.
Nonlinear reflection thus occurs at an angle $\theta = \arcsin (c k_x / \omega)$ relative to the LiF surface normal.

\paragraph*{The nonlinear polarization amplitude $\mathbf{P}$}

For the particular nonlinear processes of four-wave mixing relevant to our experiment, namely sum- and difference-frequency mixing, the amplitude $\mathbf{P}$ of the nonlinear polarization of the medium is linked to the specific third order nonlinear susceptibility tensors $\chi^{(3)}\left(-\omega_\text{X} - 2\omega_\text{I}; \omega_\text{X}, \omega_\text{I}, \omega_\text{I}\right)$ and $\chi^{(3)}\left(-\omega_\text{X} + 2\omega_\text{I}; \omega_\text{X}, -\omega_\text{I}, -\omega_\text{I}\right)$, respectively. Based on the cubic symmetry of the LiF crystal (space group $m\overline{3}m$) these tensors have 27 elements which are different from zero with only four of them being independent \cite{Butcher1990}. Skipping the dependence on the frequencies for convenience, one complete set of independent, non-zero elements is $\chi^{(3)}_{x,x,x,x}$, $\chi^{(3)}_{x,x,y,y}$, $\chi^{(3)}_{x,y,x,y}$ and $\chi^{(3)}_{x,y,y,x}$ in a Cartesian frame of reference with axes coinciding with the crystal axes \cite{Butcher1990}.

According to the experimental situation, the LiF crystal's $z$-axis is chosen to coincide with the $z$-axis of the plane of incidence of the FEL and NIR laser beams, whereas the crystal's $x$-axis enclosed an angle $\varphi$ with the $x$-axis of the plane of incidence (see Fig. \ref{geometry} of the main text). The components of the electric field amplitudes of the FEL and NIR plane waves in the nonlinear  medium may be written $\mathbf{E}_\text{X} = \left(E_{\text{X},x}, 0, E_{\text{X},z}\right)$ and $\mathbf{E}_\text{I} = \left(E_{\text{I},x}, 0, E_{\text{I},z}\right)$. This representation uses as the frame of reference the $x$- and $z$-axes of the plane of incidence together with the corresponding orthogonal $y$-axis. The $y$-components of the amplitudes are both set to zero, assuming the waves are polarized in the plane of incidence, just as the setting in the experiment. With these assumptions the components of the induced nonlinear polarization $(P_x, P_y, P_z)$ in the same reference frame can be written as
\begin{linenomath*}
	\begin{align}
		P_x =& \chi^{(3)}_{x,x,x,x}E_{\text{X},x}E_{\text{I},x}^2 \left(\sin^4\varphi + \cos^4\varphi\right) + \chi^{(3)}_{x,x,y,y}E_{\text{X},x}E_{\text{I},z}^2 + \left(\chi^{(3)}_{x,y,x,y} + \chi^{(3)}_{x,y,y,x}\right)E_{\text{X},z}E_{\text{I},x}E_{\text{I},z} \nonumber\\
		&+ \frac{1}{2}\left(\chi^{(3)}_{x,x,y,y} + \chi^{(3)}_{x,y,x,y} + \chi^{(3)}_{x,y,y,x}\right)E_{\text{X},x}E_{\text{I},x}^2 \sin^22\varphi\\
		P_y =& \frac{1}{4}\left(\chi^{(3)}_{x,x,y,y} + \chi^{(3)}_{x,y,x,y} + \chi^{(3)}_{x,y,y,x} - \chi^{(3)}_{x,x,x,x}\right)E_{\text{X},x}E_{\text{I},x}^2 \sin 4\varphi \\
		P_z =& \chi^{(3)}_{x,x,x,x}E_{\text{X},z}E_{\text{I},z}^2 + \chi^{(3)}_{x,x,y,y}E_{\text{X},z}E_{\text{I},x}^2 + \left(\chi^{(3)}_{x,y,x,y} + \chi^{(3)}_{x,x,y,y}\right)E_{\text{X},x}E_{\text{I},x}E_{\text{I},z}
	\end{align}
\end{linenomath*}
This relation supposes the nonlinear process of sum-frequency mixing. For difference-frequency mixing one has to use the complex conjugate NIR electric field strength components in the equations for the amplitude components of the nonlinear polarization above.

In our model for $\chi^{(3)}\left(-\omega_\text{X} - 2\omega_\text{I}; \omega_\text{X}, \omega_\text{I}, \omega_\text{I}\right)$ and $\chi^{(3)}\left(-\omega_\text{X} + 2\omega_\text{I}; \omega_\text{X}, -\omega_\text{I}, -\omega_\text{I}\right)$ we simplify the LiF crystal symmetry by assuming the medium to be invariant under the full rotation group. This introduces an additional constraint for the independent elements of the nonlinear susceptibility tensor above. Only three of the four elements remain independent \cite{Butcher1990}
\begin{linenomath*}
	\begin{equation}
		\chi^{(3)}_{x,x,y,y} + \chi^{(3)}_{x,y,x,y} + \chi^{(3)}_{x,y,y,x} - \chi^{(3)}_{x,x,x,x} = 0 \ .
	\end{equation}
\end{linenomath*}
As one may already expect, this relation eliminates any dependence of the induced nonlinear polarization $\mathbf{P}$ above on the angle $\varphi$.

\paragraph*{The model for the linear and 3\textsuperscript{rd} order susceptibilities}

Computing the amplitude of the electric field (Eq.~\ref{refamp}, main text) of the reflected sum- and difference-frequency waves requires the knowledge of the linear dielectric constant $\epsilon(\omega)$ of LiF in the relevant photon energy range. We constructed $\epsilon(\omega)$ using the measured linear reflection off LiF in Fig.~\ref{processes}(B). As a model a set of seven discrete, homogeneously broadened resonances was chosen to simulate the structures found in the reflection curve by suitably choosing their positions, widths and oscillator strengths. Since the measurement did not determine the reflection coefficient but only represents the intensity of the reflected light the absolute scale for the dielectric constant had to be set using a reported LiF absorption coefficient at a certain photon energy. We utilized the measured absorption coefficient at 70\,eV photon energy reported in \cite{Milgram1962}.

\begin{figure}
	\centering
	\includegraphics[width=0.75\linewidth]{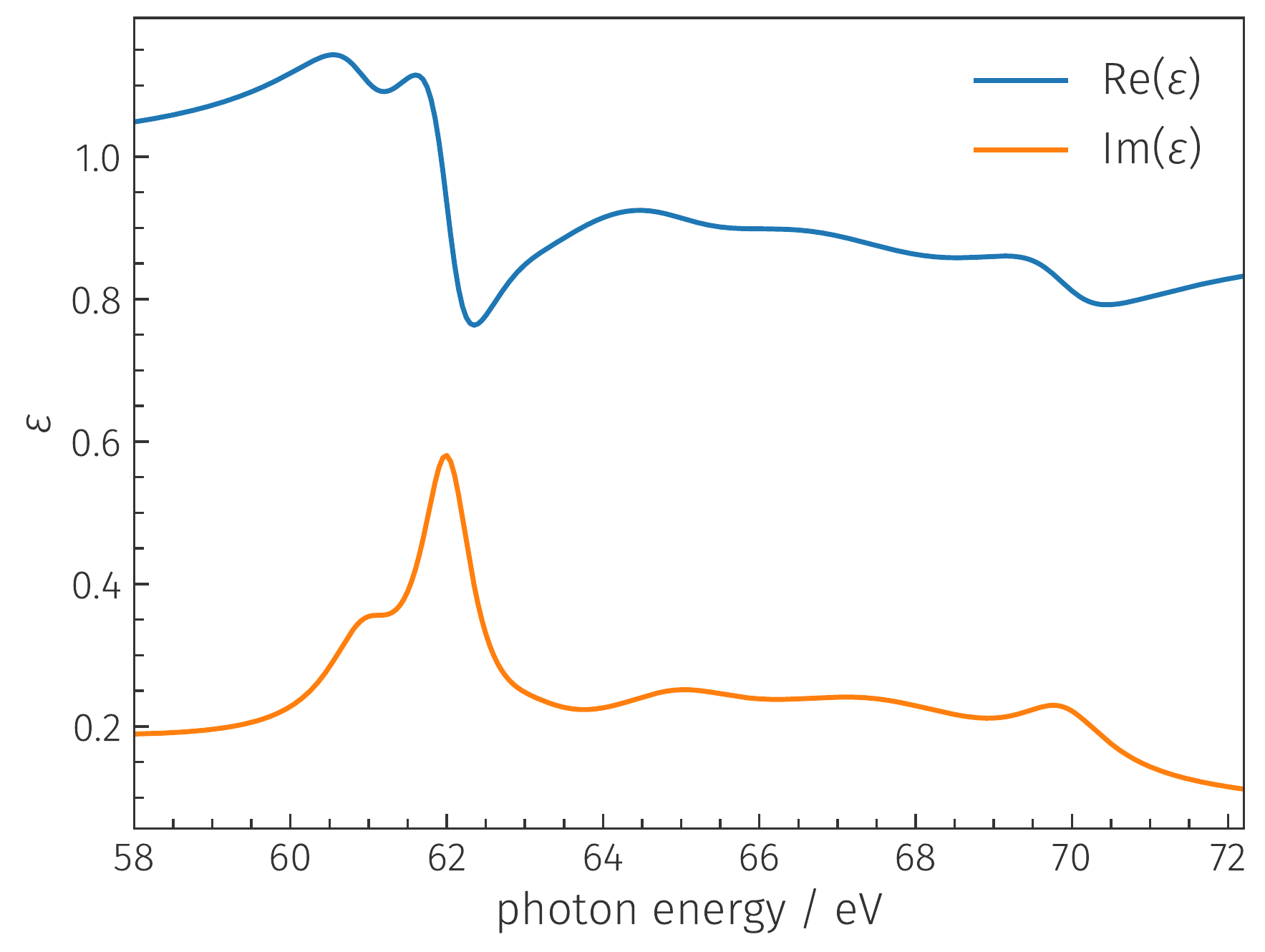}
	
	\caption{The calculated linear dielectric constant $\epsilon(\omega)$ in the photon energy range between $58\, \mathrm{eV}$ and $72\, \mathrm{eV}$. The blue line shows the real and the orange one the imaginary part of $\epsilon$.}
	\label{epsilon-calc}
\end{figure}

The relation
\begin{linenomath*}
	\begin{equation}
		\alpha \left(\omega\right) = \sum_{j} \frac{f_j}{\omega_j^2 - \omega^2 - i \gamma_j \omega}
		\label{alpha}
	\end{equation}
\end{linenomath*}
for the microscopic reaction of LiF to an applied electric field in the frequency range of interest is employed to determine the dielectric constant. It is based on molecular polarizability (see \cite{Jackson1998}). The adjustable parameters $\omega_j$, $\gamma_j$ and $f_j > 0$ are chosen so as to simulate the measured LiF reflection in Fig. \ref{processes}B. The Clausius-Massotti equation
\begin{linenomath*}
	\begin{equation}
		\alpha\left(\omega\right) = 3 \frac{\epsilon\left(\omega\right) - 1}{\epsilon\left(\omega\right) + 2}
	\end{equation}
\end{linenomath*}
links the microscopic reaction to the dielectric constant $\epsilon$ \cite{Jackson1998}. The parameters $\omega_j$, $\gamma_j$ and $f_j > 0$ which result in a reasonable fit of the experimental linear reflection off LiF (see Fig. \ref{processes}(B), the blue line) are gathered in table \ref{table}. The real and imaginary parts of the dielectric constant $\epsilon$ corresponding to this choice of parameters are shown in Fig. \ref{epsilon-calc}. The main, $p$-type LiF core exciton resonance is responsible for the maximum of the imaginary part of $\epsilon(\omega)$ at 62\,eV while its low energy shoulder represents the suspected $s$-type core exciton.

\begin{table}
	\renewcommand\thetable{S1}
	\centering
	\begin{tabular}{l|c|c|c|c|c|c|c}
		$j$ & 1 & 2 & 3 & 4 & 5 & 6 & 7 \\
		\hline
		$\omega_j$ / eV & 49.82 & 60.97 & 62.07 & 63.12 & 65.0 & 67.55 & 69.92 \\
		$\gamma_j$ / eV & 50.0 & 1.1 & 0.79 & 1.3 & 2.4 & 4.3 & 1.45 \\
		$f_j$  / $\mathrm{eV^2}$ & 42.0 & 0.59 & 1.54 & 0.17 & 0.92 & 2.77& 0.75 \\
	\end{tabular}
	\caption{The parameters entering Eq.~\ref{alpha} for the linear microscopic reaction of LiF to an applied electromagnetic wave in the photon energy range between $\approx 58\, \mathrm{eV}$ and $\approx 72\, \mathrm{eV}$ which is relevant to the experiment. The dielectric constant based on this choice of the parameters allows to reasonably reproduce the experimental LiF reflectivity as shown in Fig.~\ref{processes}(B) of the main text.}
	\label{table}
\end{table}

The calculation of the third order nonlinear susceptibility tensor is based on the expression for its components given in reference \cite{Butcher1990} (page 93). Since in the experiment a detectable, resonance like enhancement of sum- and difference-frequency mixing was observed when the LiF p-type exciton was involved, we only take this resonance into account together with the suspected $s$-type excitonic resonance in determining the dependence of $\chi^{(3)}$ on the driving FEL photon energy. In the expression for $\chi^{(3)}$ we therefore only employ the resonance positions $\omega_j$ and widths $\gamma_j$ in table \ref{table} with $j = 2, 3$ which correspond to these resonances. For the nonlinear susceptibility we neglect a dipole coupling of the $s$-type exciton to the ground electronic state. However, we take into account a potential, dipole allowed coupling of this $s$- to the $p$-type main exciton. We use this simplification since we think the main influence on sum- and difference-frequency generation by the suspected $s$-type exciton is through its coupling to the $p$-type exciton which is driven by the applied NIR laser field. The NIR photon energy of $1.55\, \mathrm{eV}$ does not much exceed the separation in energy of these two excitons ($\omega_3 - \omega_2 = 1.1\, \mathrm{eV}$ according to table \ref{table}).

With this simplification, only two dipole matrix elements are relevant in the expression for $\chi^{(3)}$: one for the dipole coupling of the $p$-type exciton to the ground state and one for the coupling of the $s$- and $p$-type excitons. For lack of information on the details of the electronic states involved we use one-electron atomic dipole matrix elements and assume the ground state is represented by a state with angular momentum $l = 0$ ($s$-state) and the excited states involved by $l = 0$ and $l = 1$ with magnetic quantum numbers $m_l = 0, \pm 1$. This approach reduces the number of free parameters in the expression for $\chi^{(3)}$ to two radial dipole matrix elements and modifies the symmetry of the LiF crystal from cubic to full rotational symmetry. Based on this simplified model we calculate the independent components of $\chi^{(3)}$ which in turn provide the amplitude of the third order nonlinear polarization amplitude $\mathbf{P}$ needed for the determination of the sum- and difference-frequency yields for comparison with the experimental result.

\paragraph*{Calibration of the MUSIX spectrometer}
For every setting of the spectrometer grating a calibration measurement was performed by varying the undulator gap of the FEL. This generated a discrete series of different FEL photon energies reflected off the LiF sample via the grating onto the CCD camera, bypassing the installed beam-block. To prevent saturation of the CCD, two 295\,nm thick zirconium attenuator foils were placed in the FEL beam path. This allowed calibrating the MUSIX spectrometer against the wavelength measurement implemented in the beamline \cite{Braune2018}. In addition, the overall consistency of the various estimated parameters used in the MUSIX spectrometer transmission calculations described below was verified by asserting an agreement between the number of photons measured by an x-ray gas-monitor-detector (XGMD) \cite{Sorokin2019} in the FEL beamline and by the MUSIX spectrometer's CCD at different wavelengths. As no linear reflectivity spectrum of the sample was acquired with the MUSIX spectrometer, a flat sample reflectivity of 0.05\,\% was used in this consistency check.

\paragraph*{Estimation of the total number of photons generated in the FWM processes}
The read out CCD camera counts were converted to an estimated number of incident photons assuming 50\,\% of these photons (a conservative estimate of the CCD's quantum efficiency) were converted into electron-hole-pairs in the silicon chip with a bandgap of 3.1\,eV. Each electron created one digital count in the analog-to-digital converter of the CCD, according to the manufacturer information for the readout frequency employed in the experimental runs. On the way from the LiF sample to the CCD the photons have been reflected off the MUSIX spectrometer gating and passed an aluminum filter foil. For the grating a 15\,\% diffraction efficiency in first order is assumed. The aluminum filter transmission (thickness 200\,nm) is retrieved from tabulated data \cite{Henke,Henke1993}. We also accounted for an estimated 12.5\,nm aluminum-oxide layer on each side of the filter. 

In a similar way, the total number of FEL photons impinging on the LiF sample per pulse was derived from pulse energy measurements using an XGMD upstream \cite{Sorokin2019}. Following the XGMD the FEL beam passed a silicon filter, three beamline mirrors, beam width limiting apertures and the incoupling mirror for the NIR laser beam which all reduced the FEL pulse energy before reaching the sample. The number of photons per pulse was determined from the total pulse energy measured by the XGMD, assuming 7\,\% of the measured pulse energy was due to FEL harmonics that were absorbed by the downstream 411\,nm thick silicon filter. Its transmission for the fundamental FEL beam was calculated from tabulated transmission data assuming a 12.5\,nm thick silicon oxide layer on each side \cite{Henke,Henke1993}. Likewise, based on tabulated data \cite{Henke,Henke1993}, the reflectivities of the three beamline mirrors (2$^\circ$ grazing angle of incidence and coated with nickel, gold and platinum, respectively) were taken into account. Wavefront-sensor measurements during beamline-alignment further suggested 25\,\% transmission through the beamline apertures and the incoupling mirror for the IR laser beam due to clipping.

Comparing the photon flux on the CCD to that measurd by the XGMD in the calibration measurements described above lends some credence to the transmission estimates made here. However, a significant uncertainty of the scaling factors involved remains. Therefore, we refer to the total number of FEL photons arriving on the LiF sample and photons generated in the FWM processes, which we show in Figs.~\ref{fig2} and \ref{fig3} of the main text, as estimates only.

\begin{figure}
	\centering
	\includegraphics[width=0.75\linewidth]{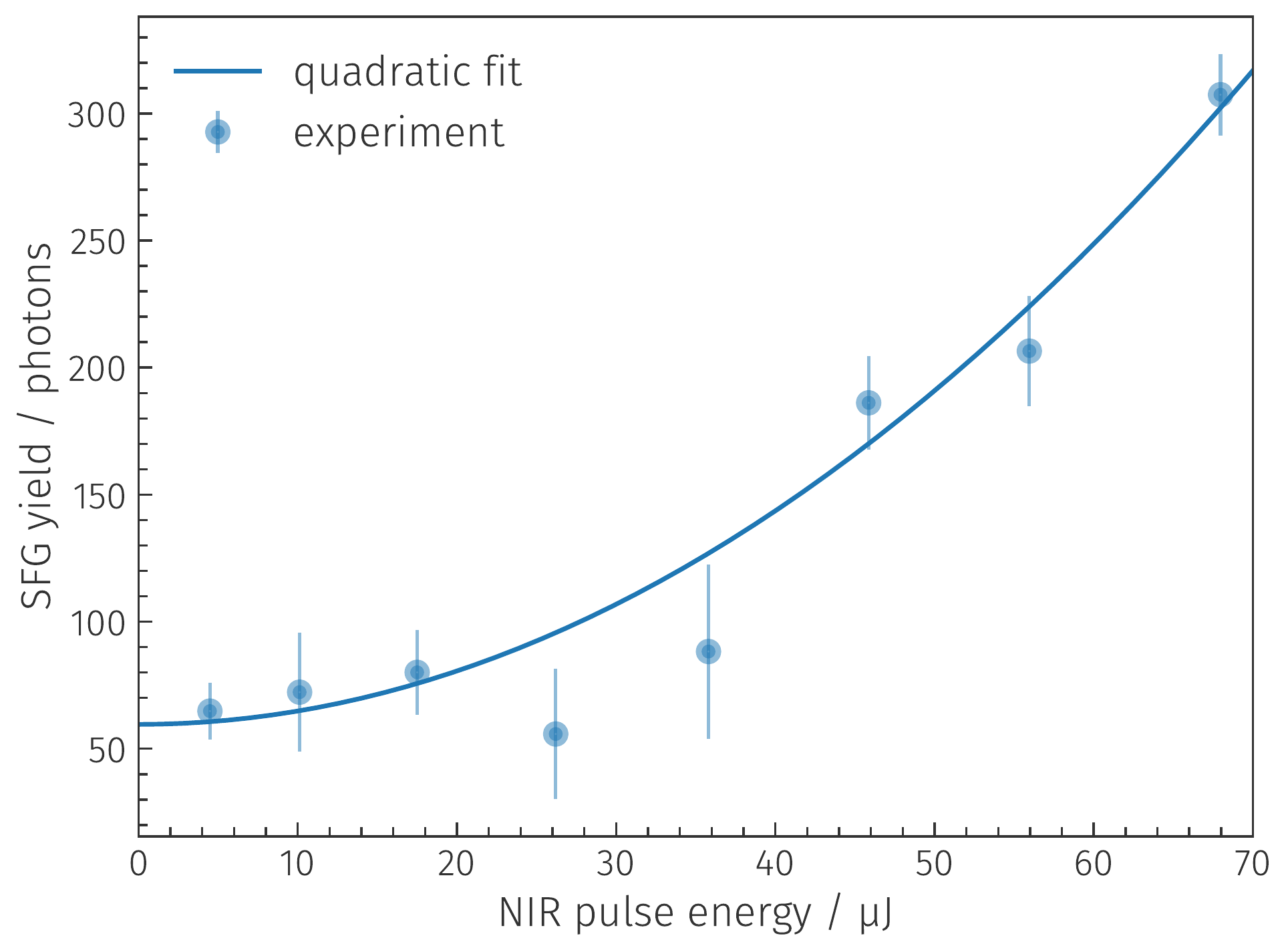}
	
	\caption{Dependence of the number of sum-frequency photons generated on the NIR laser pulse energy arriving on the LiF crystal. The FEL laser photon energy was set to 59.25\,eV, i.e. 3-photon resonant with the LiF exciton resonance ($\omega_\text{exc} = \omega_\text{X} + 2 \omega_\text{I}$). The blue line represents a fit to the data points using a quadratic polynomial in the NIR pulse energy.}
	\label{powerdependence}
\end{figure}

\paragraph*{The experimental dependence of the frequency conversion on the NIR laser pulse energy}
To further support the nature of the observed nonlinear processes we determined the dependence of the frequency conversion yield on the pulse energy of the NIR laser pulses. In the regime of low conversion, as was the case in the experiment, the dependence is expected to be quadratic. This is just what the experiment indicates as Fig. \ref{powerdependence} shows. The measurement was done with the FEL photon energy set to 59.25\,eV. The nonlinear process involved was sum-frequency mixing three-photon resonant with the LiF exciton resonance ($\omega_\text{exc} = \omega_\text{X} + 2 \omega_\text{I}$). The data points allow fitting a quadratic dependence of the sum-frequency photon yield on the NIR laser pulse energy (blue line in Fig. \ref{powerdependence}). There was a certain amount of residual stray light photons present which was impossible to eliminate by background subtraction. It is responsible for the non-zero number of photons detected with the NIR photon energy approaching zero.

\end{document}